\documentclass[aps,prd,onecolumn,showkeys,showpacs,amssymb]{revtex4}
\usepackage{epsfig}
\usepackage{graphics}
\usepackage{graphicx}
\usepackage{amsmath,amssymb}
\usepackage{amsfonts}
\usepackage{bm}
\usepackage[usenames,dvipsnames]{color}
\usepackage{amssymb}
\usepackage{psfrag}
\usepackage{times}
\usepackage[varg]{txfonts}
\usepackage{xspace}
\usepackage{float}
\usepackage{placeins} 
\usepackage{capt-of} 
\usepackage[colorlinks, pdfborder={0 0 0}]{hyperref} 
\usepackage[nameinlink,capitalise]{cleveref}

\definecolor{LinkColor}{rgb}{0.75, 0, 0}
\definecolor{CiteColor}{rgb}{0, 0.5, 0.5}
\definecolor{UrlColor}{rgb}{0, 0, 0.75}
\hypersetup{linkcolor=LinkColor}
\hypersetup{citecolor=CiteColor}
\hypersetup{urlcolor=UrlColor}
\usepackage[utf8]{inputenc}
\usepackage{ulem}
\newcommand{\Mathematica}{\texttt{Wolfram Mathematica}\xspace}

\begin{document}

\newcommand{\be}{\begin{equation}}
\newcommand{\ee}{\end{equation}}

\newcommand{\xj}[1]{\textcolor{red}{\sf{[Xisco: #1]}} }
\newcommand{\pf}[1]{\textcolor{red}{\sf{[Feola: #1]}} }
\newcommand{\ca}[1]{\textcolor{red}{\sf{[Capozziello: #1]}} }
\newcommand{\rc}[1]{\textcolor{red}{\sf{[Cianci: #1]}} }
\newcommand{\sv}[1]{\textcolor{red}{\sf{[Vignolo: #1]}} }

\title{The mass-radius relation for neutron stars in $f(R)=R+\alpha\/R^2$ gravity: a comparison between  purely metric and  torsion formulations}
\author{P. Feola$^{1,3}$, Xisco Jiménez Forteza$^{3,5}$, S. Capozziello$^{2,3,4}$, R. Cianci$^{1}$, S.
Vignolo$^{1}$ }

\affiliation{$^{1}$DIME Sez. Metodi e Modelli Matematici,
Universit\`a di Genova,  Via All' Opera Pia 15a - 16100 Genova
(Italy)}

\affiliation{$^{2}$ Dipartimento di Fisica, ``E. Pancini''
Universit\`{a} ``Federico II'' di Napoli, 
Compl. Univ. Monte S. Angelo Ed. G, Via Cinthia, I-80126
Napoli (Italy)}

\affiliation{$^{3}$ INFN Sez. di Napoli, Compl. Univ. 
Monte S. Angelo Ed. G, Via Cinthia, I-80126 Napoli (Italy)}

\affiliation{$^{4}$Laboratory for Theoretical Cosmology,
Tomsk State University of Control Systems and Radioelectronics (TUSUR), 634050 Tomsk, Russia.}

\affiliation{$^{5}$Dipartimento di Fisica, "Sapienza"
Universit\`a di Roma,  Piazzale Aldo Moro 5, 00185, Roma
(Italy)}

\begin{abstract}
Within the framework of $f(R)=R+\alpha\/R^2$ gravity, we study realistic models of neutron stars, using  equations of state  compatible with the LIGO constraints. i.e. APR4, MPA1, SLy, and WW1. By numerically solving modified Tolman-Oppenheimer-Volkoff  equations, we investigate the Mass--Radius relation in both   metric and  torsional $f(R)=R+\alpha\/R^2$ gravity models. In particular, we observe that  torsion effects  decrease the compactness and total mass of neutron star with respect to the General Relativity predictions, therefore mimicking the effects of a repulsive massive field. The opposite occurs in the metric theory, where mass and compactness increase with $\alpha$, thus inducing an excess of mass that overtakes the standard General Relativity limit. We also find that the sign of $\alpha$ must be reversed whether one considers the metric theory (positive) or torsion (negative) to avoid blowing up solutions. This could draw an easy test to either confirm or discard one or the other theory by determining  the sign of  parameter $\alpha$.
\end{abstract}


\date{\today}

\maketitle
\section{Introduction}
Compact objects, such as Neutron Stars (NS), are astrophysical objects that can  be described by  General Relativity (GR).
These relativistic stars are natural laboratories for studying
the behavior of  high-density nuclear matter using an appropriate equation of state (EoS), which relates the pressure and density of degenerate matter. This allows one to obtain the Mass-Radius relation,  $\mathcal{M-R}$, 
and other macroscopic properties such as the tidal deformability  and the stellar momentum of inertia \cite{Steiner:2014pda}.

Since the internal structure of a NS cannot be reproduced in the laboratory because of the extreme conditions in which it operates, only theoretical models can be formulated where there are a very large number of EoS candidates.
The astrophysical measurements of the macroscopic properties of  NS are very useful because they allow us to understand what can be  realistic EoS . In fact, they can provide information on whether the EoS is soft or stiff and what is the pressure several times the density of nuclear saturation \cite{Lattimer:2015nhk, Hebeler:2013nza, Ozel:2016oaf, Steiner:2017vmg}. Therefore, measuring the mass value of a NS could help us to describe matter at extreme gravity regimes.

Einstein's theory describe accurately the physical properties that govern the stability of NS where Chandrasekhar, considering degenerate matter, fixed a theoretical upper limit of $1.44 M_\odot$ so that the stability of a non-rotating degenerate star is conserved \cite{Chandrasekhar:1931ih}. Instead, as confirmed by several astrophysical observations, there exist binary systems with NS  having  mass values that violates this limit allowing larger masses \cite{Barziv:2001ad,Rawls:2011jw,Mullally:2009rr,Nice:2005fi,2010Natur.467.1081D,Bao}.

To study these observational evidences, as already done in some previous works, developed in   metric formalism, \cite{capquark,Astashenok:2013vza,Astashenok:2014pua,Astashenok:2014gda,Astashenok:2014nua,Capozziello:2015yza},  Extended Theories of Gravity \cite{Capozziello:2011et, dagostino} can be used.  In particular   $f(R)$ gravity, i.e.  a class of Lagrangians considering   a generic function of the Ricci curvature scalar. The primary objective is to obtain the $\mathcal{M-R}$   relation for a NS that allows, given an EoS, to derive the maximum mass value.

From a cosmological point of view, $f(R)$ theories, beside addressing  in a straightforward way the inflationary paradigm \cite{Starobinsky:1980te},   could be useful in view of problems like the accelerated expansion of the universe (the $dark$  $energy$ issue),  confirmed by several observations   \cite{Perlmutter:1998np,Riess:1998cb, Riess:2004nr,Spergel:2006hy,Schimd:2006pa,McDonald:2004eu}, and the problem of the formation of large-scale structures, called $dark$  $matter$. 
Unlike the standard Concordance Lambda Cold Dark Matter ($\Lambda$CDM) Model \cite{Bahcall:1999xn,Bamba,Joyce}, similar results can be obtained without considering dark components but extending the gravitational sector at infrared scales \cite{Capozziello:2002rd, Capozziello:2003tk, Nojiri:2003ft, Carroll:2003wy, Olmo:2011uz, Nojiri:2010wj, Capozziello:2010zz, Capozziello:2011et, delaCruzDombriz:2012xy}.  Specifically, $f(R)$ gravity
is acquiring a  growing interest because it allows a good description of  gravitating structures without non-baryonic dark matter:   extra degrees of freedom of gravitational field can be dealt as  effective scalar fields contributing to the structure formation and stability \cite{Capozziello:2012ie,Cembranos:2008gj}.   In this perspective, it is possible  to unify the cosmic acceleration \cite{Capozziello:2002rd,delaCruzDombriz:2006fj},  the early-time inflation \cite{Starobinsky:1980te,Planck:2013jfk}, thus leading to a complete picture of the evolution of the Universe 
\cite{Nojiri:2003ft,Ferrara,Sebastiani,Bamba2,Bamba3,Nojiri3,Elizalde} and large-scale structures therein \cite{delaCruzDombriz:2008cp,Abebe,Abebe2,Oikonomou}.
However, the \textit{dark side}  and the $f(R)$ descriptions are, in some sense, equivalent at large scale so one needs an \textit{experimentum crucis} capable of discriminating among the two competing pictures. Discovering new particles out of the Standard Model or addressing gravitational phenomena that escape from the GR description could be an approach to fix this challenging issue.  Observing exotic stars  modeled by some alternative theory of gravity could be a goal in this perspective. 

In this paper,  we derive the $\mathcal{M-R}$ diagram, using realistic EoS compatible with the LIGO constraints \cite{TheLIGOScientific:2017qsa} for a $f(R)=R+\alpha R^2$ Lagrangian, using two different approaches: the purely metric theory  and a theory with torsion that allows one to introduce the spin degrees of freedom in GR \cite{Hehl:1976kj}. In our specific model, the torsion field is due to the non-linearity of $f(R)$. Here the mass-energy is the source of curvature and the spin is the source of torsion. In this way, torsion contributions could provide additional information for compact stars in extreme gravity regimes.

The goal of this paper is to obtain realistic $\mathcal{M-R}$ relation by solving numerically  a modified system of equations, derived from  Tolman-Oppenheimer-Volkoff (TOV)  \cite{Oppenheimer:1939ne}  equations, and compare results with the LIGO constraints. Specifically,  we shall consider  quadratic corrections to the Ricci scalar and  discuss   models with and without torsion comparing them with GR.

The  paper is organized as  follows. In Section II we derive the TOV equations for $f(R)$ gravity in the metric and torsion formalism.  Section III is devoted to  the problems related to the numerical aspects  of TOV equations in  $f(R)$ gravity. In Section IV we derive the numerical  solutions of stellar structure equations and  compare the results of the $\mathcal{M-R}$ relations. Discussion and conclusions  are given in Sec. V.
\section{Tolman-Oppenheimer-Volkov equations in $f(R)$ gravity}
\label{sec:me}
\subsection{The  metric theory}
\label{subsec:metric}
In the  metric formulation, 
the action for $f(R)$ gravity (in units for $G=c=1$) is given by 
\begin{equation}\label{action}
{\cal A}=\frac{1}{16\pi }\int d^4x \sqrt{-g}[f(R) + {\cal L}_{{\rm matter}}],
\end{equation}
where $f(R)$ is a function of the scalar curvature $R$, $g$ is determinant of the metric tensor $g_{ij}$ and ${\cal L}_{\rm matter}$ 
is the matter Lagrangian.
Varying the action \eqref{action} with respect to the metric tensor $g_{ij}$, one gets  the
field equations:
\begin{equation}\label{field}
f'(R)R_{ij}-\frac{1}{2}f(R) g_{ij}-(\nabla
_{i}\nabla _{j}-g_{ij}\Box )f'(R)=8 \pi  \Sigma_{ij}.
\end{equation}
In eqs. \eqref{field}, $R_{ij}$ is the Ricci tensor, $f'(R)$ denotes the derivative of $f(R)$ with
respect to the scalar curvature, 
$\displaystyle{\Sigma_{ij}= \frac{-2}{\sqrt{-g}}\frac{\delta\left(\sqrt{-g}{\cal L}_m\right)}{\delta g^{ij}}}$
is the energy--momentum tensor of matter and  $\square=\frac{1}{\sqrt{-g}}\frac{\partial}{\partial x^{j}}\left(\sqrt{-g} g^{ij} \frac{\partial}{\partial x^{i}}\right)$ indicates the  covariant d'Alembert operator. Here we  adopt the signature  $\left(+,-,-,-\right)$. 

In order to describe stellar objects, we assume that the metric is static and 
spherically symmetric of the form:
\begin{equation}\label{metric}
    ds^2= e^{2\psi} dt^2 -e^{2\lambda}dr^2 -r^2 (d\theta^2 +\sin^2\theta d\phi^2),
\end{equation}
where $\psi$ and $\lambda$ are functions depending only on the radial coordinate $r$. We  assume that the interior of  the star matter is described by a perfect fluid,  with energy--momentum tensor $\Sigma_{ij}=\mbox{diag}(e^{2\psi}\rho, e^{2\lambda}p, r^2 p, r^{2}p\sin^{2}\theta)$,  where $\rho =\rho(r)$ and $p=p(r)$ are the matter density and pressure respectively.

By a direct calculation, it is possible to show that field eqs. (\ref{field}), evaluated in the metric \eqref{metric}, are equivalent to a set of equations consisting of
the Tolmann-Oppenheimer-Volkov (TOV) equations for $f(R)$ gravity and a  continuity equation given by the contracted Bianchi identity $\nabla^i \Sigma_{ij}=0$. Specifically, 
the TOV equations for $f(R)$gravity are  
\begin{eqnarray}
\label{dlambda_dr}
\frac{d\lambda}{dr}&=&\frac{ e^{2 \lambda }[r^2(16 \pi \rho + f(R))-f'(R)(r^2 R+2)]+2R_{r}^2 f'''(R)r^2+2r f''(R)[r R_{r,r} +2R_{r}]+2 f'(R)}{2r \left[2 f'(R)+r R_{r} f''(R)\right]},
\end{eqnarray}
\begin{eqnarray}\label{psi1}
\frac{d\psi}{dr}&=&\frac{ e^{2 \lambda }[r^2(16 \pi p -f(R))+ f'(R)(r^2 R+2)]-2(2rf''(R)R_{r}+ f'(R))}{2r \left[2 f'(R)+r R_{r} f''(R)\right]},
\end{eqnarray}
while the continuity equation is  
\begin{equation}\label{hydro}
    \frac{dp}{dr}=-(\rho +p)\, \frac{d\psi}{dr} .
\end{equation}
Here $R_{r}$ and $R_{r,r}$ denote respectively the first and second derivative of $R(r)$ with respect to radial coordinate $r$.  
In order to solve numerically the equations \eqref{dlambda_dr}, \eqref{psi1} and \eqref{hydro}, we can consider  the scalar curvature $R$ as an independent dynamical field. In doing this, we need an additional equation which is directly obtained form the definition of scalar curvature:
\begin{equation}\label{Ricci0}
R=2e^{-2\lambda}\left[\psi_r^2 - \psi_r\lambda_r + \psi_{r,r} + \frac{2\psi_r}{r} - \frac{2\lambda_r}{r} + \frac{1}{r^2} - \frac{e^{2\lambda}}{r^2}\right],
\end{equation}
Indeed, inserting the content of eqs. \eqref{dlambda_dr}, \eqref{psi1} and \eqref{hydro} into \eqref{Ricci0}, we get the dynamical equation for $R$:
\begin{equation}\label{TOVR}
\frac{d^2R}{dr^2}=R_{r}\left(\lambda_{r}+\frac{1}{r}\right)+\frac{f'(R)}{f''(R)}\left[\frac{1}{r}\left(3\psi_{r}-\lambda_{r}+\frac{2}{r}\right)- e^{2 \lambda }\left(\frac{R}{2} + \frac{2}{r^2}\right)\right]- \frac{R_{r}^2f'''(R)}{f''(R)}.
\end{equation}
Finally, the numerical solution of the resulting dynamical equations relies on the assignment of a suitable EoS, $p=p({\rho})$, relating pressure and density inside the star, as well as of initial data (i.e. values of the fields at the center of the star).

\subsection{The theory with torsion}
\label{subsec:torsion}
In $f(R)$ gravity with torsion, 
the gravitational and dynamical fields are pairs ($g$, $\Gamma$) consisting of a pseudo-Riemannian metric $g$ and a metric compatible linear connection $\Gamma.$ with non--vanishing torsion.
  
The corresponding field equations are obtained by varying the action functional (\ref{action}) independently  with respect to the metric and the connection.  It is worth noticing that now ${R}$ refers to the scalar curvature associated with the dynamical connection $\Gamma$.

Moreover, we recall that any metric compatible linear connection $\Gamma$ may be decomposed as the sum   
\begin{equation}\label{conn}
\Gamma_{ij}^{\;\;\;h} = \tilde{\Gamma}_{ij}^{\;\;\;h} - K_{ij}^{\;\;\;h},
\end{equation}
where $\tilde{\Gamma}_{ij}^{\;\;\;h}$ is the Levi--Civita connection associated with the given metric $g$ and $K_{ij}^{\;\;\;h}$ denotes the contorsion tensor, related to the torsion tensor $T_{ij}^{\;\;\;h}= \Gamma_{ij}^{\;\;\;h} - \Gamma_{ji}^{\;\;\;h}$ by the relation
\cite{Hehl:1971qi}:
\begin{equation}\label{2.3}
K_{ij}^{\;\;\;h} = \frac{1}{2}\/\left( - T_{ij}^{\;\;\;h} + T_{j\;\;\;i}^{\;\;h} - T^h_{\;\;ij}\right).
\end{equation}
The contorsion tensor \eqref{2.3} verifies the antisymmetry property $K_{i}^{\;\;j\;h} = - K_{i}^{\;\;h\;j}$ and, together with the metric tensor $g$, identifies the actual degrees of freedom of the theory.

Making use of eqs. \eqref{conn} and \eqref{2.3}, we can decompose the  Ricci and the scalar curvature of the dynamical connection respectively as:
\begin{equation}\label{Ricci:dyn}
R_{ij}=\tilde{R}_{ij} + \tilde{\nabla}_jK_{hi}^{\;\;\;h} - \tilde{\nabla}_hK_{ji}^{\;\;\;h} + K_{ji}^{\;\;\;p}K_{hp}^{\;\;\;h} - K_{hi}^{\;\;\;p}K_{jp}^{\;\;\;h}
\end{equation}
and 
\begin{equation}\label{defR}
R=\tilde{R} + \tilde{\nabla}_jK_{h}^{\;\;jh} - \tilde{\nabla}_hK_{j}^{\;\;jh} + K_{j}^{\;\;jp}K_{hp}^{\;\;\;h} -
K_{h}^{\;\;jp}K_{jp}^{\;\;\;h} 
\end{equation}
where $\tilde{R}_{ij}$ and $\tilde{R}$ are the Ricci and  the scalar curvature of the Levi--Civita connection induced by the metric $g$.

In the absence of matter spin density, variations of (\ref{action}) yield the field equations \cite{Capozziello:2007tj,Capozziello:2008yx,Capozziello:2008kb,Capozziello:2010iy,Capozziello:2010}:
\begin{equation}\label{2.1ax}
f'\/(R)R_{ij} -\frac{1}{2}f\/(R)g_{ij}=8 \pi  \Sigma_{ij},
\end{equation}
and
\begin{equation}\label{2.1b}
T_{ij}^{\;\;\;h}
=\frac{1}{2f'(R)}\frac{\partial f'(R)}{\partial x^{p}}(\delta^{p}_{j}\delta^{h}_{i}-\delta^{p}_{i}\delta^{h}_{j}),
\end{equation}
where $\Sigma_{ij}$ denotes again the energy-momentum tensor of matter, and the non--linearity of the gravitational Lagrangian function $f(R)$ becomes a source of torsion.

Now,  by  inserting  eqs. \eqref{Ricci:dyn} and \eqref{2.1b}  into  eqs. (\ref{2.1ax}),  
it is possible to show that the whole set of field equations evaluated in the metric \eqref{metric} is equivalent to the system formed by the following two TOV equations  
\begin{eqnarray}
\label{dlambda_drt}
\frac{d\lambda}{dr}&=&\frac{e^{2 \lambda }\left[r^2(16 \pi   \rho + f(R))-f'(R)(r^2 R+2)\right]+2R_{r}^2 f'''(R)r^2+2r^2 f''(R)\left[ R_{r,r} +\frac{2R_{r}}{r}-\frac{3f''(R) R_{r}^2}{4f'(R)}\right] +2 f'(R)}{2r \left[2 f'(R)+r R_{r} f''(R)\right]},
\end{eqnarray}

\begin{eqnarray}\label{psi1t}
\frac{d\psi}{dr}&=&\frac{ e^{2 \lambda }[r^2(16 \pi p - f(R))+f'(R)(r^2 R +2)]-2rf''(R)R_{r}\left[2+\frac{3f''(R) r R_{r}}{4f'(R)}\right]-2 f'(R)}{2r \left[2 f'(R)+r R_{r} f''(R)\right]},
\end{eqnarray}
together with the continuity equation
\begin{eqnarray}\label{eqcont2}
\frac{dp}{dr}=-(\rho +p)\, \frac{d\psi}{dr}, 
\end{eqnarray}
which also holds in the present case \cite{CV1,CV2}.

Also in the torsional case, we consider the scalar curvature $R$ as an independent dynamical variable, introducing a consequent additional equation derived from the very definition of $R$ itself. In fact, inserting eqs. (\ref{2.3}) and (\ref{2.1b}) into \eqref{defR}, evaluating all in the metric \eqref{metric} and making use of eqs. (\ref{dlambda_drt}) and (\ref{psi1t}), we obtain the evolution equation:
\begin{equation}\label{TOVRT}  
\frac{d^2R}{dr^2}=R_{r}\left(\lambda_{r}+\frac{1}{r}\right)-\frac{2f'(R)}{f''(R)}\left[\frac{1}{r}\left(3\psi_{r}-\lambda_{r}+\frac{2}{r}\right)-e^{2 \lambda }\left(\frac{R}{2} + \frac{2}{r^2}\right)\right] -R_{r}^2\left(\frac{f'''(R)}{f''(R)}+\frac{3f''(R)}{2f'(R)}+\frac{3\psi_{r}}{R_{r}}+\frac{9}{rR_{r}}\right),
\end{equation}
Again, in order to be solved, the set of dynamical equations \eqref{dlambda_drt}, \eqref{psi1t}, \eqref{eqcont2} and \eqref{TOVRT}  for the unknowns $R, \lambda,\psi, p$ and $ \rho$ must be completed by an EoS and initial data.
\subsection{The  $f(R)=R+\alpha R^2$ model}
\label{sect_quadratic}
\noindent
We consider here the specific form of $f(R)$:
\begin{equation}
f(R)=R+\alpha R^2,
\label{fr_form_quadratic}
\end{equation}
where $\alpha$ is the coupling parameter of the quadratic curvature correction. This model is specially suitable to account for cosmological inflation, where higher-order curvature terms naturally lead to cosmic accelerated expansion. The quadratic term emerges in strong gravity regimes, while at Solar System scales and, more in general, in the weak field regime, the linear term predominates. 

This statement can be easily demonstrated because any analytic $f(R)$ model, in the weak field limit, presents  a 
Yukawa-like correction in the gravitational potential except for $f(R)= R$ where only the Newtonian potential is recovered. As shown in \cite{New1,New2}, such a correction is  relevant at very large scales (e.g. at galactic scales and beyond \cite{Capozziello:2012ie}) with respect to Solar System and does not affect classical experimental  constraints of GR. As a consequence, $R^2$ terms are relevant only in the strong field regime.

 Since the interior of a NS could present  energy conditions in some sense similar to those  early universe  \cite{Astashenok:2014pua},  the model \eqref{fr_form_quadratic}
is particularly suitable for our considerations. In this model, eqs. (\ref{dlambda_dr}), (\ref{psi1}) and (\ref{TOVR}) take the explicit form:
\begin{eqnarray}
\label{dlambda_dr alfa}
\frac{d\lambda}{dr}&=&\frac{ e^{2 \lambda }[16 \pi r^2 \rho - 2-\alpha R(r^2 R+4)]+4\alpha(r^2 R_{r,r} +2 r R_{r}+R)+2}{4r \left[1+\alpha(2R+r R_{r})\right]},
\end{eqnarray}
\begin{eqnarray}\label{psi1 alfa}
\frac{d\psi}{dr}&=&\frac{ e^{2 \lambda }[16 \pi r^2 p + 2+\alpha R(r^2 R+4)]-4\alpha(2rR_{r}+R)-2}{4r \left[1+\alpha(2R+r R_{r})\right]},
\end{eqnarray}
\begin{equation}\label{TOVR alfa}
\frac{d^2R}{dr^2}=R_{r}\left(\lambda_{r}+\frac{1}{r}\right)+\frac{1+2\alpha R}{2\alpha}\left[\frac{1}{r}\left(3\psi_{r}-\lambda_{r}+\frac{2}{r}\right)- e^{2 \lambda }\left(\frac{R}{2} + \frac{2}{r^2}\right)\right],
\end{equation}
while eqs. (\ref{dlambda_drt}), (\ref{psi1t}) and (\ref{TOVRT}) become respectively:
\begin{eqnarray}
\label{dlambda_dr alfa torsion}
\frac{d\lambda}{dr}&=&\frac{ e^{2 \lambda }[16 \pi r^2 \rho - 2-\alpha R(r^2 R+4)]+4\alpha \left[r^2 R_{r,r} +2 r R_{r}+R-\frac{3\alpha r^2 R_{r}^2}{2(1+2\alpha R)}\right] +2}{4r \left[1+\alpha(2R+r R_{r})\right]},
\end{eqnarray}
\begin{eqnarray}\label{psi1 alfa torsion}
\frac{d\psi}{dr}&=&\frac{ e^{2 \lambda }[16 \pi r^2 p + 2+\alpha R(r^2 R+4)]-4\alpha \left[2rR_{r}+R+\frac{3\alpha r^2 R_{r}^2}{2(1+2\alpha R)}\right]-2}{4r \left[1+\alpha(2R+r R_{r})\right]},
\end{eqnarray}
\begin{equation}\label{TOVR alfa torsion}
\frac{d^2R}{dr^2}=R_{r}\left(\lambda_{r}+\frac{1}{r}\right)-\frac{1+2\alpha R}{\alpha}\left[\frac{1}{r}\left(3\psi_{r}-\lambda_{r}+\frac{2}{r}\right)- e^{2 \lambda }\left(\frac{R}{2} + \frac{2}{r^2}\right)\right]-R_{r}^2\left(\frac{3\alpha}{1+2\alpha R}+\frac{3\psi_{r}}{R_{r}}+\frac{9}{rR_{r}}\right).
\end{equation}
Clearly the torsion contributions emerge in the second system. 
In the next Section, we shall discuss numerical solutions for the interior space--time of spherically symmetric NS in both metric and torsional $f(R)=R+\alpha R^2$ gravity. Our aim is to compare  the solutions of the above two systems of differential  equations in order to point out the torsion contribution with respect to the purely metric one.

In view of this, it is worth noticing that, \textit{in vacuo}, $f(R)=R+\alpha\/R^2$ gravity with torsion amounts to GR \cite{Capozziello:2007tj,Capozziello:2010}. Therefore, under the assumption of spherical symmetry, in the case with torsion, the space--time outside the star has to   coincide with the Schwarzschild one. In order to compare  the two models, it is then reasonable and consistent assuming the external Schwarzschild solution as the space--time outside the star also in the case of purely metric theory. It is worth noticing  that the Schwarzschild external solution  is actually a vacuum solution for purely metric $f(R)=R+\alpha\/R^2 $ gravity as demonstrated in \cite{Whitt, Mignemi}.

Therefore, viable  interior solutions, at the boundary,  have to match  the external Schwarzschild solution. In this regard, we recall that junction conditions for $f(R)$ gravity have been studied in \cite{Deruelle} for the purely metric formulation, and in \cite{VCC,CV1,CV2} for the theory with torsion. Referring the reader to \cite{Deruelle, VCC} for more details, we assume the following   junction conditions at the stellar radius 
\begin{equation}\label{junction_metric}
\lambda \in C^0, \quad \psi \in C^1, \quad R \in C^1 \quad\quad {\rm in  \; purely \; metric \; case}\,,
\end{equation} 
\begin{equation}\label{junction_torsion}
\lambda \in C^0, \quad \psi \in C^1, \quad \frac{dR}{dr} \in C^0 \quad\quad {\rm in \;  torsional \; case}\,.
\end{equation}
Outside the star $\lambda$, $\psi$ and $R$ refer to the corresponding Schwarzschild quantities. Eqs. \eqref{junction_metric} and \eqref{junction_torsion} are the conditions at the stellar radius to be satisfied by the numerical solutions we shall investigate in the next Sections.

\section{Numerical aspects of the TOV equations in $f(R)=R+\alpha\/R^2$ gravity}
\label{num}

The TOV equations presented in Sec.  \ref{sec:me}, together with an EoS, form a closed system of equations that can be solved numerically once a suitable set of initial conditions are provided. The EoS accounts for the behavior of the matter fields in the NS at nuclear level.  However, it  also dominates the NS macroscopic properties as the total mass $\cal M$, radius $\mathcal{R}_S$ and compactness $\cal{C}=\cal{M}/\mathcal{R}_S$. The total mass $\cal M$ and the radius $\mathcal {R}_S$ may vary significantly depending on the state of  matter in the NS interior where $C \approx [0.02,0.25]$, being $C=0.5$ the black hole solution. On the other hand, the knowledge of the macroscopic properties provides a direct insight to understand the particle interactions, energy transport and state of the matter in the NS core. Until recently, there were placed only vague constraints on the EoS of NSs from electromagnetic observations  \cite{Radice:2017lry}. The recent LIGO-Virgo binary neutron star (BNS) observation has significantly clarified the state of  art concerning the EoS physics. The largest accuracy of the gravitational wave (GW) channel in relation to the electromagnetic (EM) observations  allowed to rule out stiffer solutions (less compact) thus reducing significantly the number of astrophysically relevant EoS. In this section, we discuss some aspects of the numerical solution of  TOV equations in the metric and torsional $f(R)$, formulations described above, for four EoS compatible with the recent LIGO constraints: APR4, MPA1, SLy, WFF1  \cite{Alford:2004pf,Mueller:1996pm,Douchin:2001sv,Wiringa:1988tp}, accurately described the piecewise polytropic fits provided in \cite{Read:2009yp}.

Then, to solve numerically the TOV equations, we use a dimensionless version of the them by re-scaling our physical variables as
\begin{equation}
r \rightarrow r/r_g, \qquad R \rightarrow R/r_g^2,\qquad p\rightarrow P/P_0, \qquad\rho \rightarrow \rho/\rho_0 \,,
\end{equation}
where
\begin{equation}
r_g=G M_{\odot}/c^2, \qquad P_0=M_{\odot}c^2/r_g^3, \qquad\rho_0=M_{\odot}/r_g^3 \,,
\end{equation}
and $M_{\odot}$ is the mass of the sun, $r_g$ is the gravitational radius ($ \simeq1.5 km$), $G$ Newton's Gravitational constant and $c$ the speed of light. 
The two systems of differential equations shown in Subsection \ref{sect_quadratic} take the following form,
\begin{align}
\label{eq:nsch}
p'=f_1(\rho,p,\psi',r),\quad
\lambda'=f_2(\lambda,R,R',R'',\rho,r),\quad
\psi'=f_3(\lambda,R,R',p,r),\quad
R''=f_4 (\lambda,\lambda',\psi',R,R',\rho,r),\quad
p=f_5(\rho)\,,
\end{align}
where the \textit{primed} variables denote radial derivatives. Therefore, we are left to setup five initial conditions (ICs) for the variables $\left \lbrace p(0), \lambda(0), \psi(0), R(0),R'(0) \right \rbrace$ to complete the numerical scheme. ICs are chosen at the center of the star $r=0$ in order to preserve regularity, thus preventing the generation of large gradients that may lead to numerical instabilities. Mathematically, this involves that any expansion around the NS center must have a zero first derivative. In particular, the scalar curvature at the NS center may be expanded as,
\begin{equation}
\label{eq:Rp0}
R(r \to 0 )\approx R(0) + R'(0) r + \frac{1}{2} R''(0)  r^2\,,
\end{equation}
where regularity involves $R'(0)=0$. Pressure and density at the center $\rho(0)=\rho_c$ and $p(0)=p_c$ are given by the EoS so they only depend on the type of fluid under consideration. For the metric potential $\lambda$, it  is natural to fix $\lambda(0)=0$, analogously to what happens in Newtonian gravity, where the $\lambda(r)$ and $\psi(r)$ variables are matched to the $m(r)$ mass of the system by,
\begin{equation}
\label{eq:lmass}
e^{2 \lambda(r)}=\left(1-2 \frac{m(r)}{r}\right)^{-1}\,,\qquad e^{2 \psi(r)} =\left(1-2\frac{m(r)}{r}\right).
\end{equation}

Notice that the variable $\psi(r)$ does not enter directly in our system of differential equations which implies that $\psi(0)$ can be defined up to any arbitrary constant. Therefore we adjust $\psi(0)$ conveniently to match (i)  the internal solutions with the external Schwarzschild solution at the stellar radius $\mathcal{R}_S$ and (ii) to obtain asymptotically the $\mathcal{O}(r^{-1})$ profile as,
\begin{equation}
\label{eq:ICas}
\lambda(r \to \infty) \approx \frac{M}{r} ,\qquad \psi(r \to \infty) \approx -\frac{M}{r} \qquad \mbox{and} \qquad  \rho (r \to \infty)=0 ,\qquad p(r  \to \infty)=0\,.
\end{equation}
The star radius is ideally defined where the pressure $p(\mathcal{R}_S)\approx 0$ though, in practice, and for numerical reasons, it is sufficient to set a ground value $\epsilon$ as $p(\mathcal{R}_S)/p_c \leq \epsilon \sim 10^{-10}$.

The fulfillment of eqs. (\ref{eq:ICas}) requires to find an optimal choice for the Ricci scalar $R_c=R(0)$. In general, this is achieved by shooting the  central value $R_c$ within some sufficiently large range  $[R_{c}^{min},R_{c}^{max}]$, containing the true value $R_c$. Then $R_c$ is found by applying bijection root-finding methods until eqs. (\ref{eq:ICas}) are satisfied up to numerical tolerance. Unfortunately, the existence of such  $R_c$ strongly depends on the particular form of the $f(R)$ model, giving rise to ghosts in case of ill-defined configuration of the model parameters. This is true for both metric and torsional $(R+\alpha\/R^2)$ theories discussed in  this work. Then we choose the sign of $\alpha$ to be the one that better matches the junction conditions at the surface of the star (\ref{junction_metric}), (\ref{junction_torsion}) for the metric and torsional theory respectively. As we evince in the following sections, the only choices that reproduce not blowing up solutions are $\alpha>0$ for the metric case and $\alpha <0$ for the torsion one. Unfortunately, these choices generate some typical tachyonic oscillations due to a bad behaved $f''(R)$ and that we could not remove numerically. This effect was also reported in \cite{Resco:2016upv} and it  shows an oscillatory behavior, in the form of a damped-sinusoid outside the star, even in the minimally perturbed scenario with $\alpha \ll 1$. These oscillations grow as the value of $\alpha$ increases and they are as well propagated to our metric potentials $\lambda(r)$ and $\psi(r)$. This inserts some ambiguity in defining the asymptotic conditions (\ref{eq:ICas}) for  large $r$ since the oscillations are not totally vanished when the numerical noise begins to dominate the solution (for $r \sim 100$). To overcome this issue and to reduce the amplitude of the oscillations, we restrict our analysis to small values of  $\alpha \in [0.001,0.1]$. As this is anyway consistent with current observational tests, in doing so, we are not discarding any relevant astrophysical scenario and this fact  allows us to set $R_c \approx R_{GR}$. This hypothesis is shown to have a minimal impact in the $\mathcal{M-R}$ diagrams as we will discuss throughout next sections. Moreover, the assumption of a Schwarzschild-type solution outside the star allows us to smooth out these oscillations and to recover a  good fulfillment of the junction conditions. According to the above positions, we justify the choice of $\alpha>0$ for the metric theory and  $\alpha<0$ for the torsional one.

Finally, the two systems of ordinary differential equations (ODE)  are solved by using a $8th$-order Runge-Kutta with adaptive step-size and high-stiffness control methods implemented in the \Mathematica package \cite{Mathematica}. These methods regulate the discretization step-size by estimating the error of the Runge-Kutta method point by point ensuring the numerical convergence of the solution step by step. The stiffness control methods use polynomial extrapolation on the short regimes where the gradients become too large. We have found these methods essential to ensure the accuracy of the solutions in the torsional formulation.
\section{Numerical solutions}
\label{sec:res}
\vspace*{-\baselineskip}
We compute the $\mathcal{M-R}$ diagrams for metric and torsional  formulations of $f(R)=R+\alpha\/R^2$ gravity. 
Due to the numerical limitations found throghout our analysis, we restrict  $|\alpha| \in [0,0.1]$ where $\alpha$ is required to be positive for the purely metric theory and negative in the theory with torsion to avoid blowing up solutions \cite{Resco:2016upv}. These values are anyway consistent with solar system tests of GR \cite{Resco:2016upv,Lombriser:2011zw}. Such tests fix light constraints on the form $f(R)\lesssim 10^{-6}$ rather than on the parameter $\alpha$, thus being translated as $ R+ |\alpha| R^2 \lesssim 10^{-6}$. Bearing in mind that curvatures themselves are expected to be small,  this leaves the parameter $\alpha$ rather unconstrained. Other tests as  E\"{o}t-Wash laboratory experiment set $\alpha \lesssim 10^{-10} m^2$.
On the contrary, there exist alternative observational space-based constraints coming from the Gravity Probe B experiment \cite{Everitt:2011hp} or the observation of the binary pulsar PSR J0737-3039 \cite{Breton:2008xy,Naf:2010zy} that set $\alpha \lesssim [5 \times 10^{11}, 2.3 \times 10^{15}] m^2$. Therefore, the discrepancies among the several experiments do not set tight bounds on the value of $\alpha$, and our choice seems to be compatible with existing data. 
\begin{figure}[!htb]
 \includegraphics[width=0.48\columnwidth]{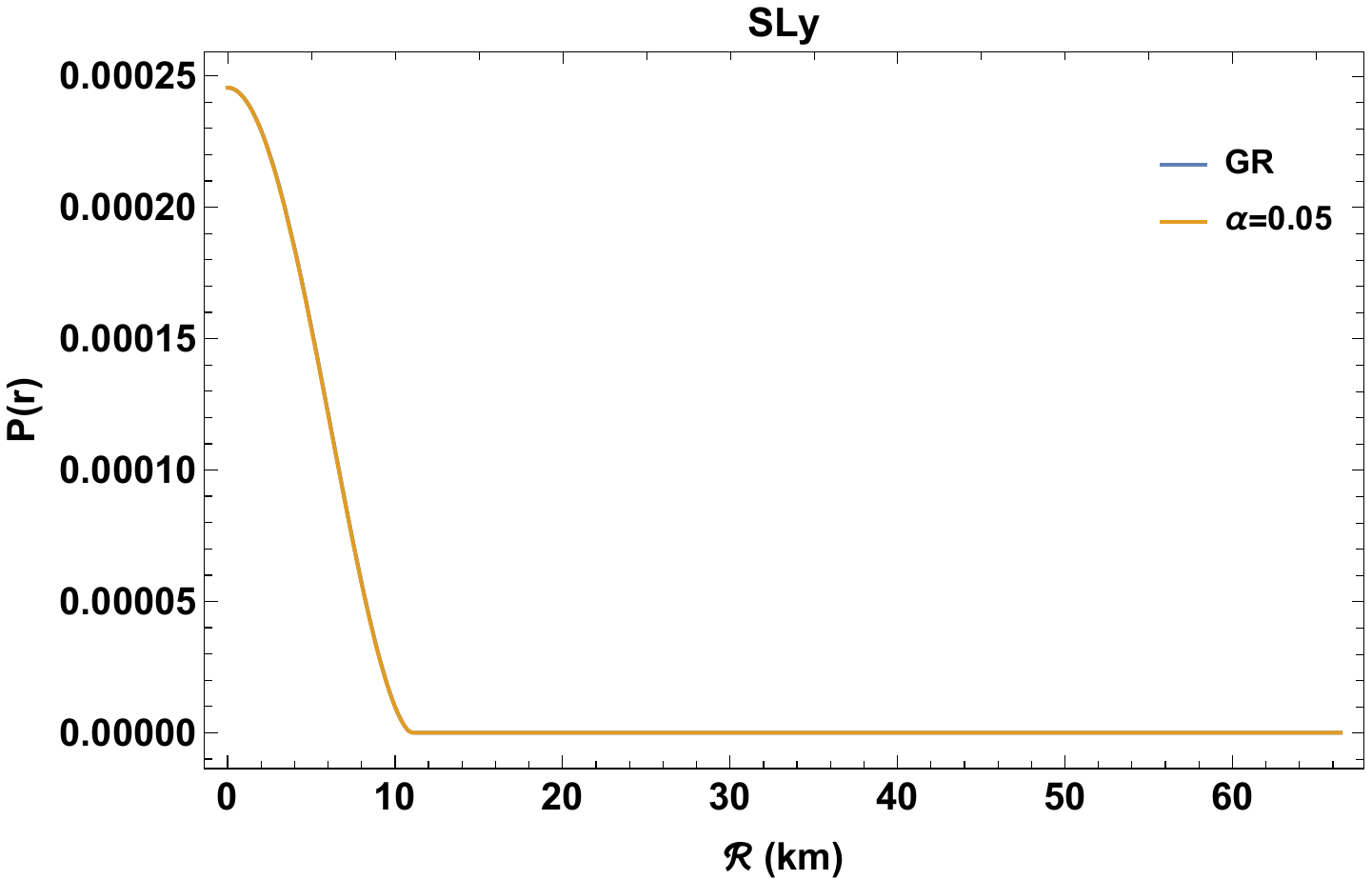}
  \includegraphics[width=0.48\columnwidth]{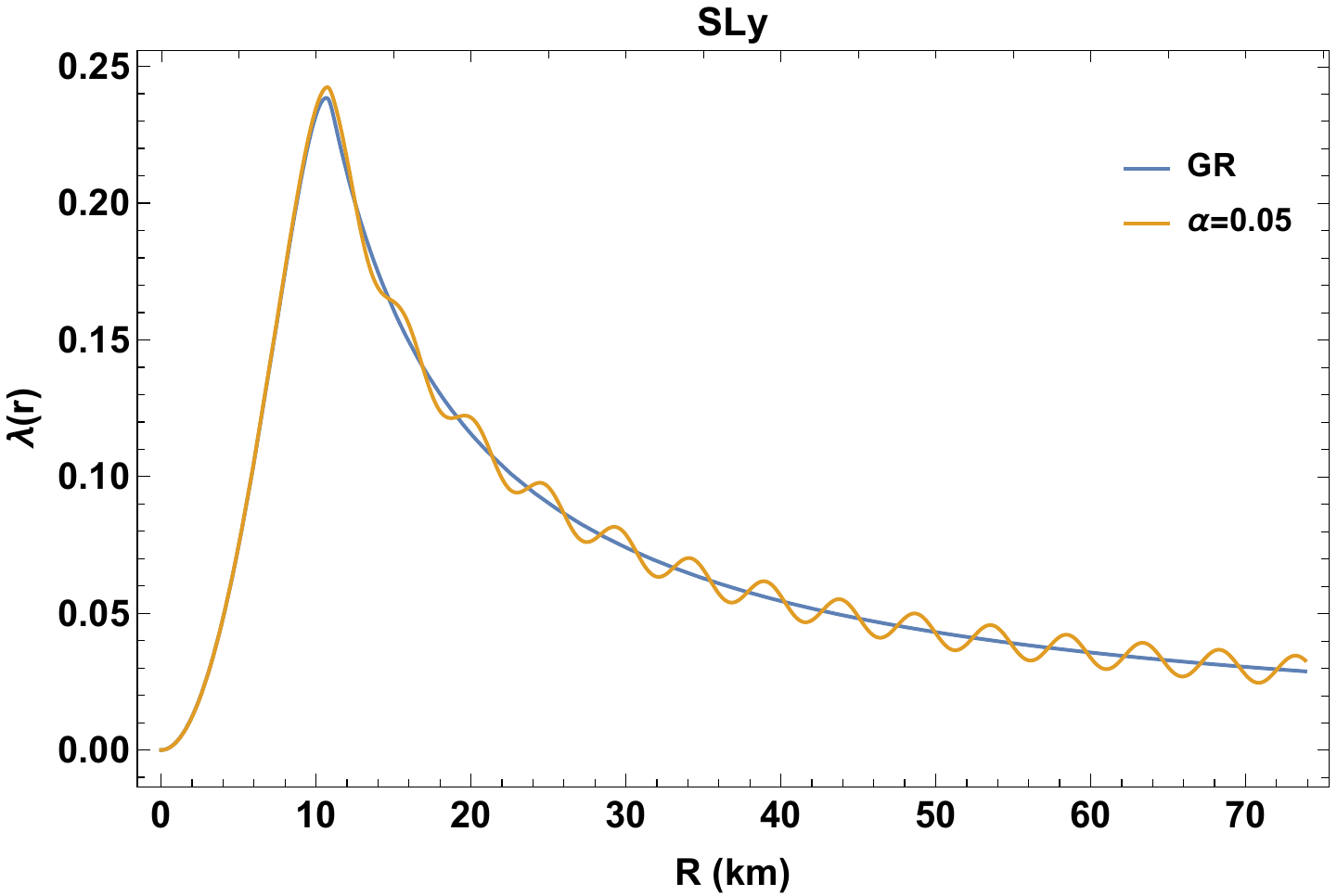}
  \includegraphics[width=0.48\columnwidth]{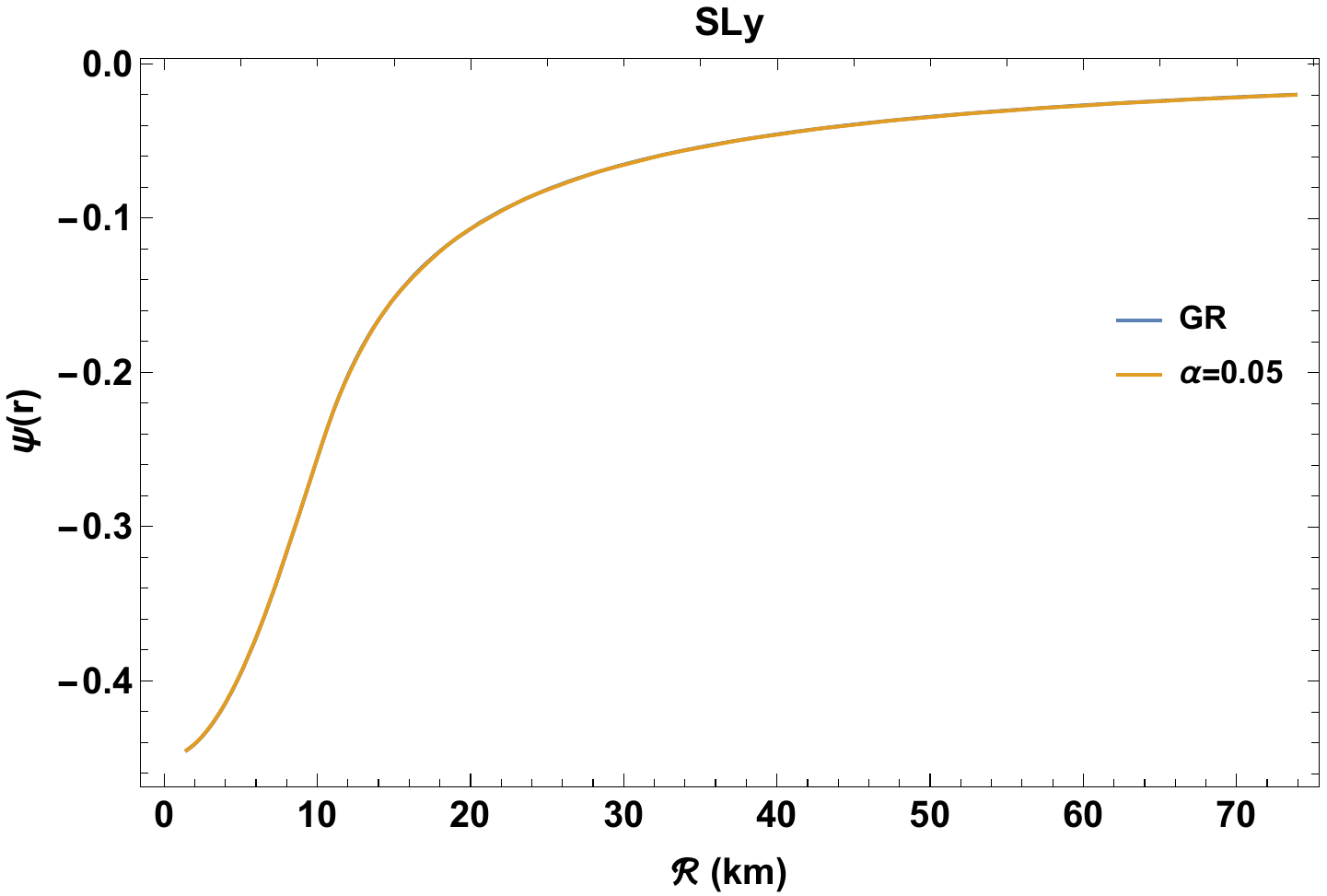}
 \includegraphics[width=0.48\columnwidth]{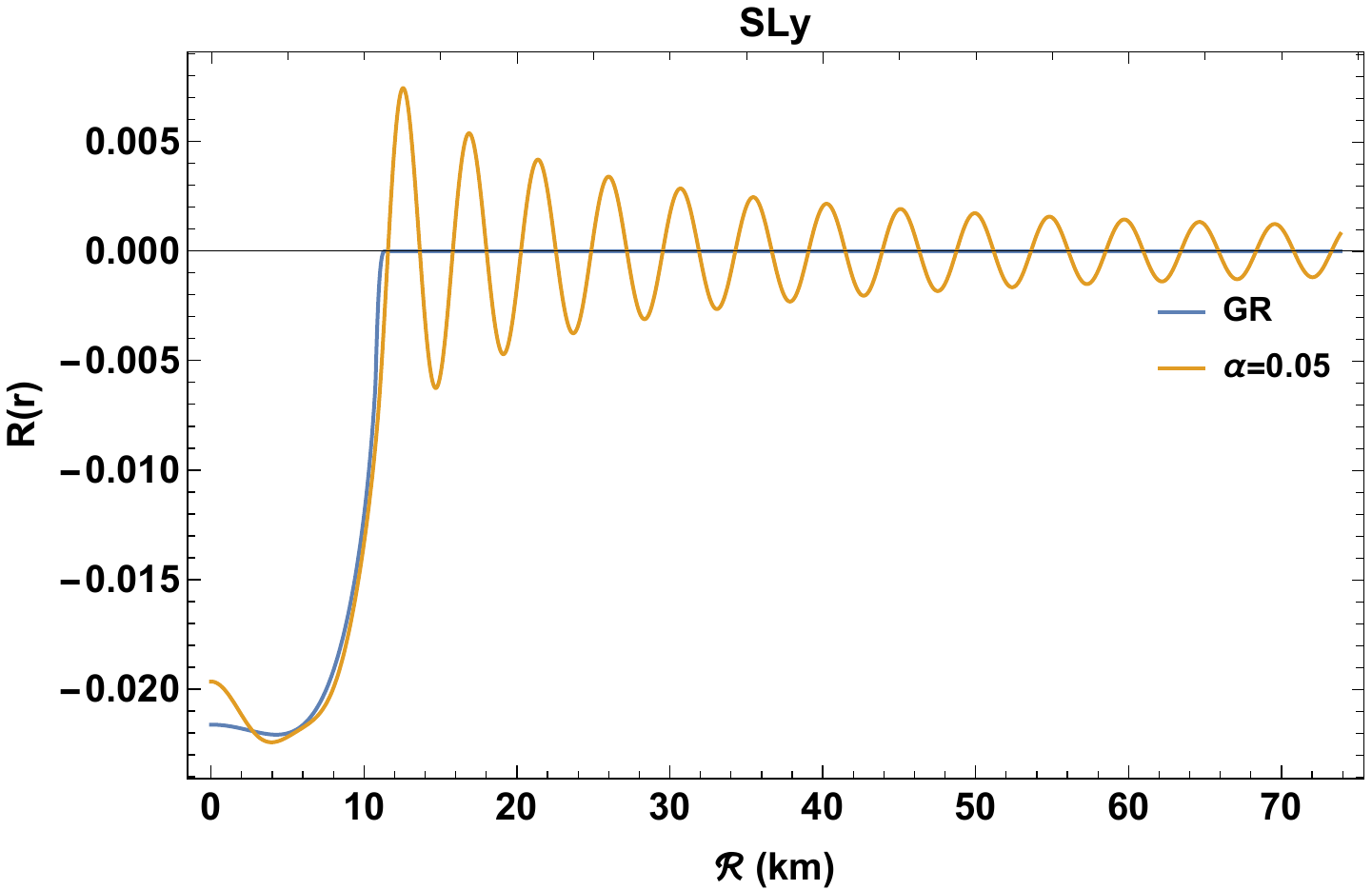}
 \caption{Solutions of the TOV equations for GR (blue) and purely metric $R+\alpha\/R^2$ with $\alpha=0.05$ (orange), using the SLy EoS. All the plotted quantities show small deviations with respect to GR. Note the asymptotic decay of the metric potentials $\lambda$ and $\psi$ as $r\rightarrow \infty$. Our choice of $\alpha$ explains the  oscillatory behavior as  reported in \cite{Resco:2016upv}.
  \label{fig:rodep}}
\end{figure}
  \vspace*{-\baselineskip}
\subsection{Purely metric theory}
 \vspace*{-\baselineskip}
The solutions of the TOV equations for the purely metric $f(R)=R+\alpha\/R^2$ model are illustrated in Figure \ref{fig:rodep}. The pressure at the center of the star $p_c$ drops quickly until it eventually gets equal to zero, thus defining the radius of the star $\mathcal{R}_S$. This radius is used as our reference point to compute the total mass $\mathcal M$ by means of eq. \eqref{eq:lmass}. 
\begin{figure}[!htb]
 \includegraphics[width=\columnwidth]{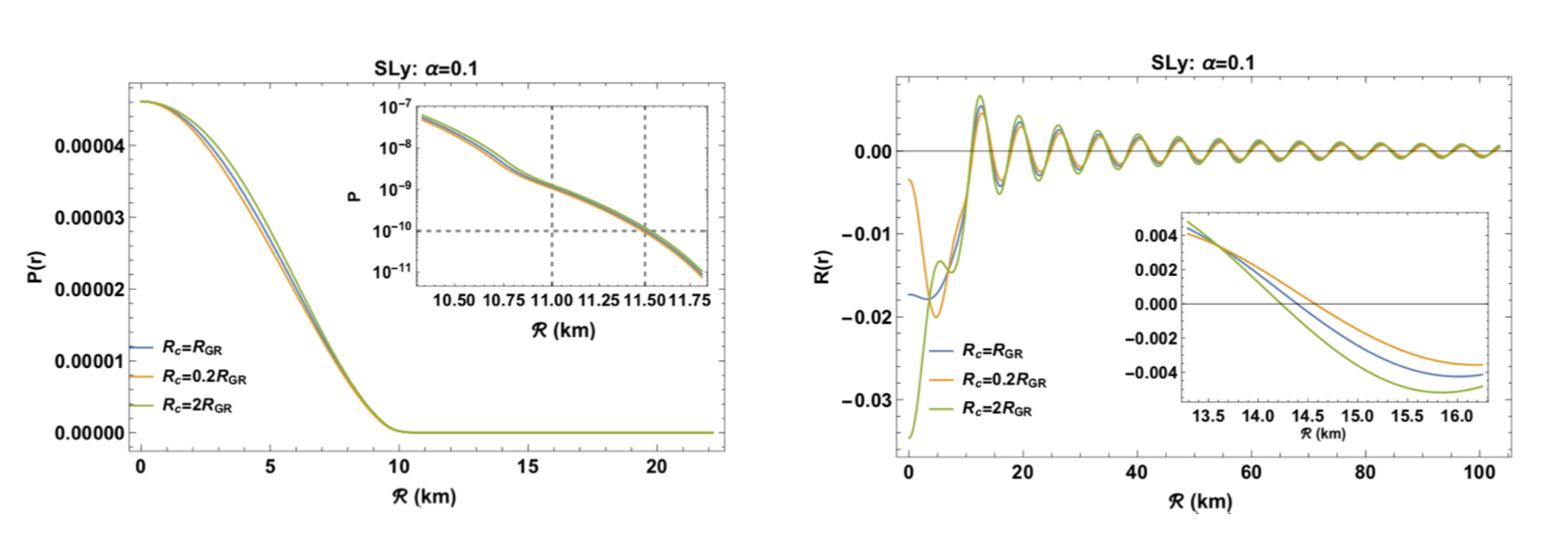}
 \vspace{-2\baselineskip}
 \caption{Profiles for the pressure $P$ (left) and the Ricci scalar $R$ (right) corresponding to ${R_c=\left\lbrace R_{GR_c},0.2R_{GR_c},2R_{GR_c}\right\rbrace}$ for the $f(R)=R+\alpha R^2$ model with $\alpha=0.1$. In the zoomed-in plot for the pressure, the grid lines fix two possible values for the radius of the star $\mathcal R_S$ that depend on to the accuracy chosen in defining its position as: $p(	\mathcal{R_S})/pc \leq \left\lbrace 10^{-9}, 10^{-10}\right\rbrace$ providing a relative difference of about $4\%$. Complementary, on the right hand side plot we show the $R=0$ point for different choices of the central value $R_c$. Notice that on the latter the effects of choosing one or another $R_c$ contribute in total about the $\sim 2\%$ between $0.2R_{GR_c}$ and $2R_{GR_c}$ choices thus this error being smaller than the our error estimate in defining $\mathcal{R}_S$.}
  \label{fig:R0dep}
\end{figure}
The numerical system exhibits some dissipative oscillations about the Ricci scalar $R$ and the metric potential $\lambda$. These oscillations naturally arise from the harmonic-form of the Ricci scalar $R(r)$ equation in vacuum \cite{Resco:2016upv}, for a non optimal choice of the Ricci scalar $R_c$ at the center of the star, and where optimal choice is here defined as that matching the Schwarzschild junction conditions at the stellar radius. Unfortunately, such a choice becomes increasingly difficult as $\alpha$ tends to zero since the system of equations become also stiffer \cite{Doneva:2013rha}. Generally speaking, this may appear to be counterintuitive, since $\alpha \to 0$ should exactly recover the GR space-time. However the asymptotic approach to $\alpha \to 0$ of the Ricci scalar equations (\ref{TOVR alfa})(\ref{TOVR alfa torsion}) are ill-defined. This is clear if, for instance, one re-expresses (\ref{TOVR alfa}) as,
\begin{equation}
\label{eq:r2}
R''=-\frac{e^{2 \lambda } \left(8 \pi  (\rho -3 p)+R\right)}{6 \alpha }-R'
   \left(-\lambda '+\psi'+\frac{2}{r}\right)\,.
\end{equation} 
Notice that the numerator of the first term is exactly zero  in GR and that ideally approaches to zero faster than  linear order in $\alpha$. However, this is not so exact when dealing with numerical uncertainties, where the same factor may behave as a $\sim 0/0$ solution for $\alpha<<1$ thus requiring much more precision on the estimation of central value $R_c$.
To overcome this issue, we have set $R(0)=R_{GR}=8\pi (3p_c -\rho_c)$ to the GR value. Though this seems apparently an arbitrary choice, we notice that, for $\alpha \lesssim<1$,  the solution must be close to  GR so the value cannot be further to that of GR. This is self-evident from Fig. \ref{fig:R0dep}, where, in  the right plot, we illustrate the variations on the pressure $p(r)$ and the Ricci scalar $R(r)$ for different choices of the central value ${R_c=\left\lbrace R_{GR_c},0.2R_{GR_c},2R_{GR_c}\right\rbrace}$. Then, notice than  the effect of varying $R_{c}$ on the radius $\mathcal{R}$ for such small values of $\alpha$ is about $\sim 2\%$ considering the maximum and minimum choices of $R_c$. This variation is then compared with the uncertainty arising from the definition of the star radius $\mathcal{R_S}$ to be the place where the pressure drops by a factor $\epsilon$. Then, in the left plot, we show that the impact of relaxing  this value to $\epsilon\sim 10^{-9}$ would generate an uncertainty of about  $4\%$, thus larger than the one from varying $R_c$. 
\begin{figure}[!htb]
 \includegraphics[width=0.48\columnwidth]{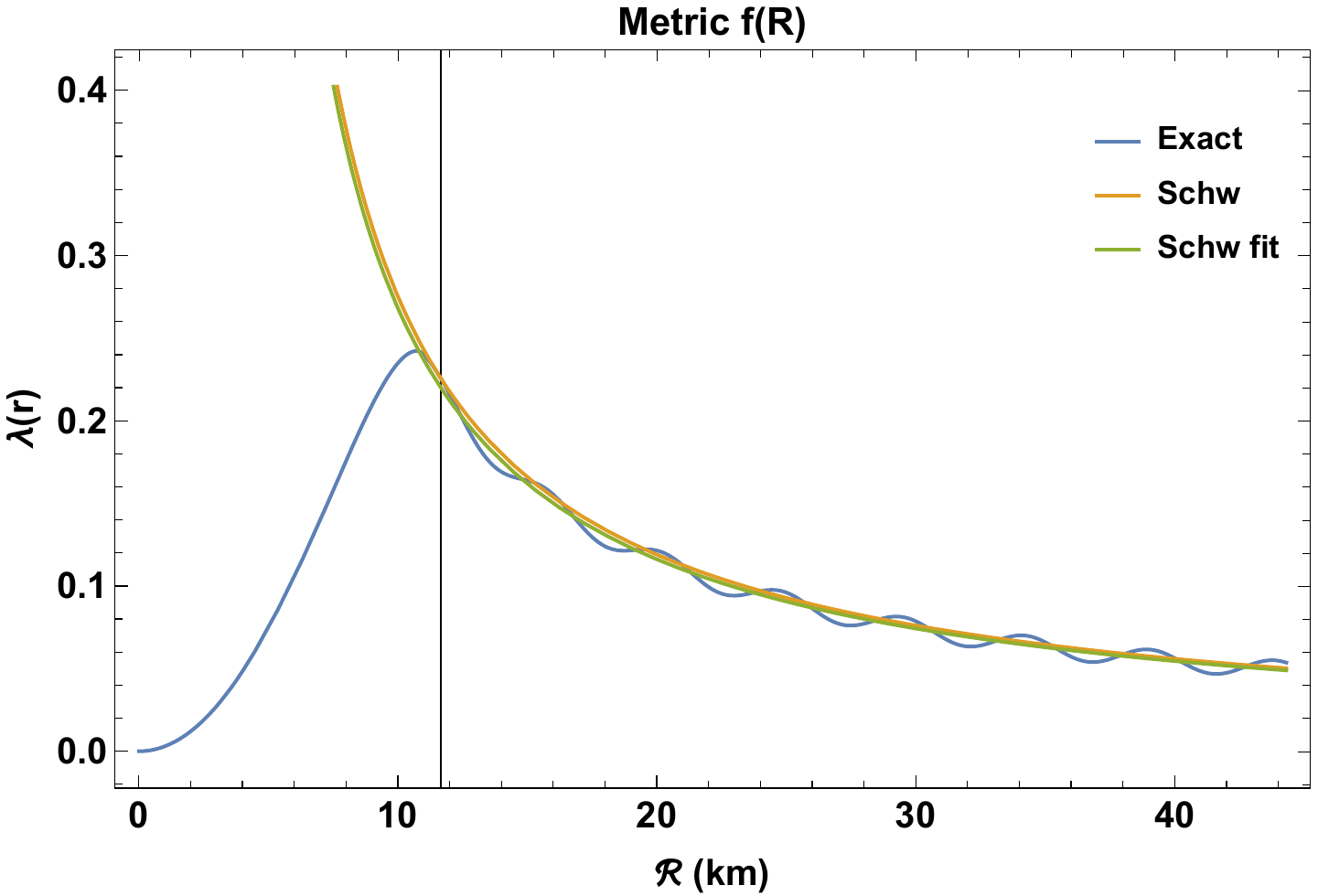}
  \includegraphics[width=0.48\columnwidth]{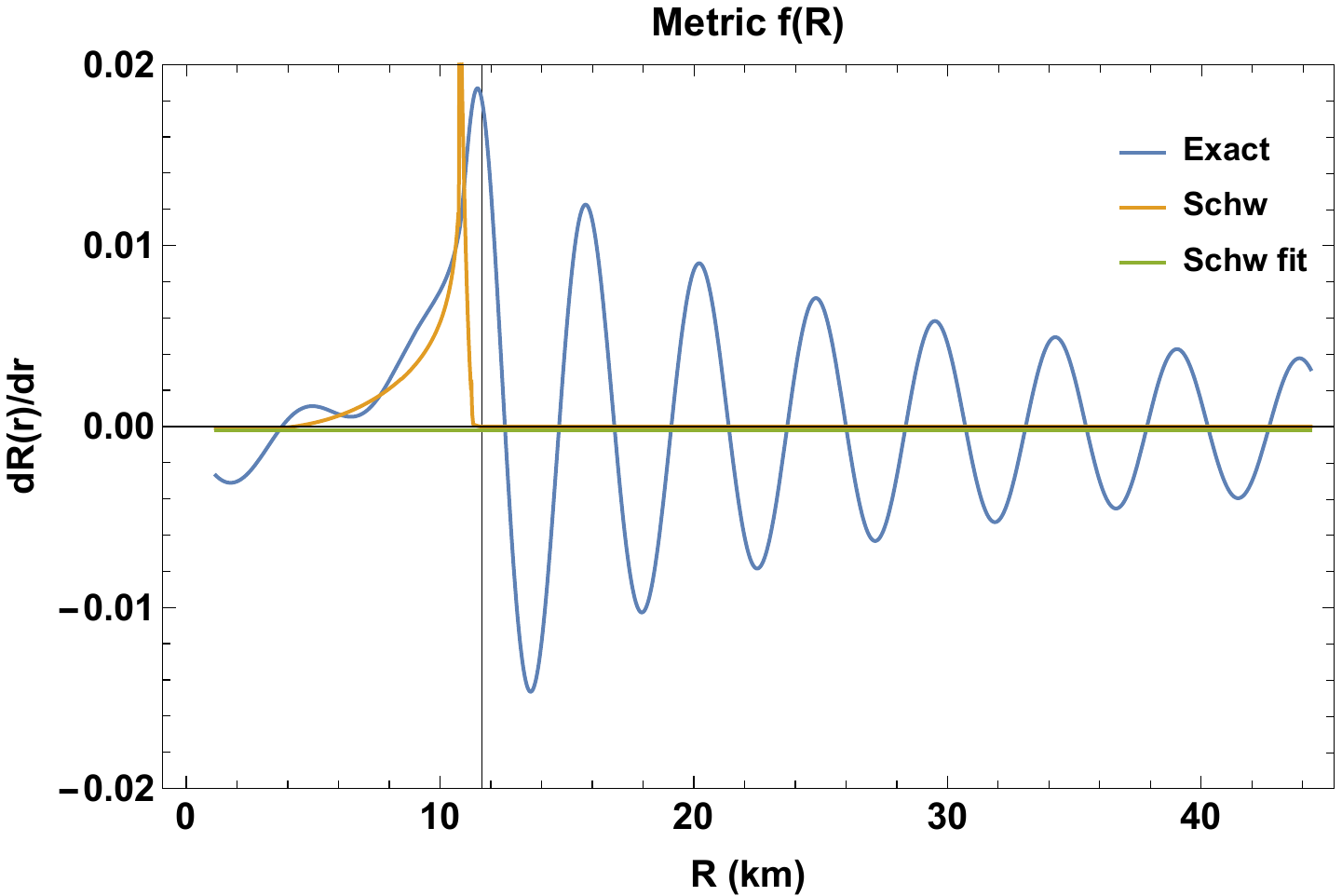}
  \includegraphics[width=0.48\columnwidth]{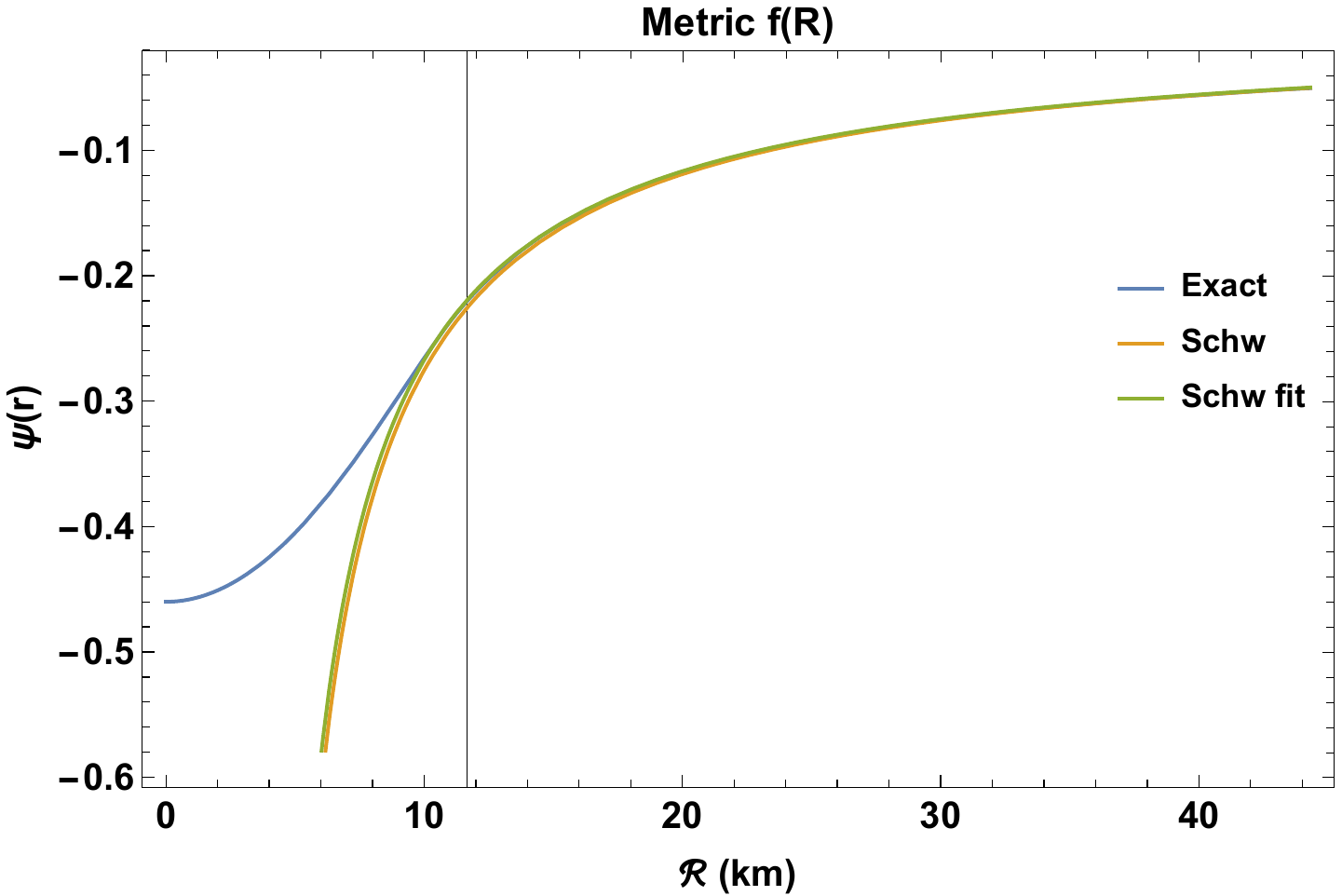}
 \includegraphics[width=0.48\columnwidth]{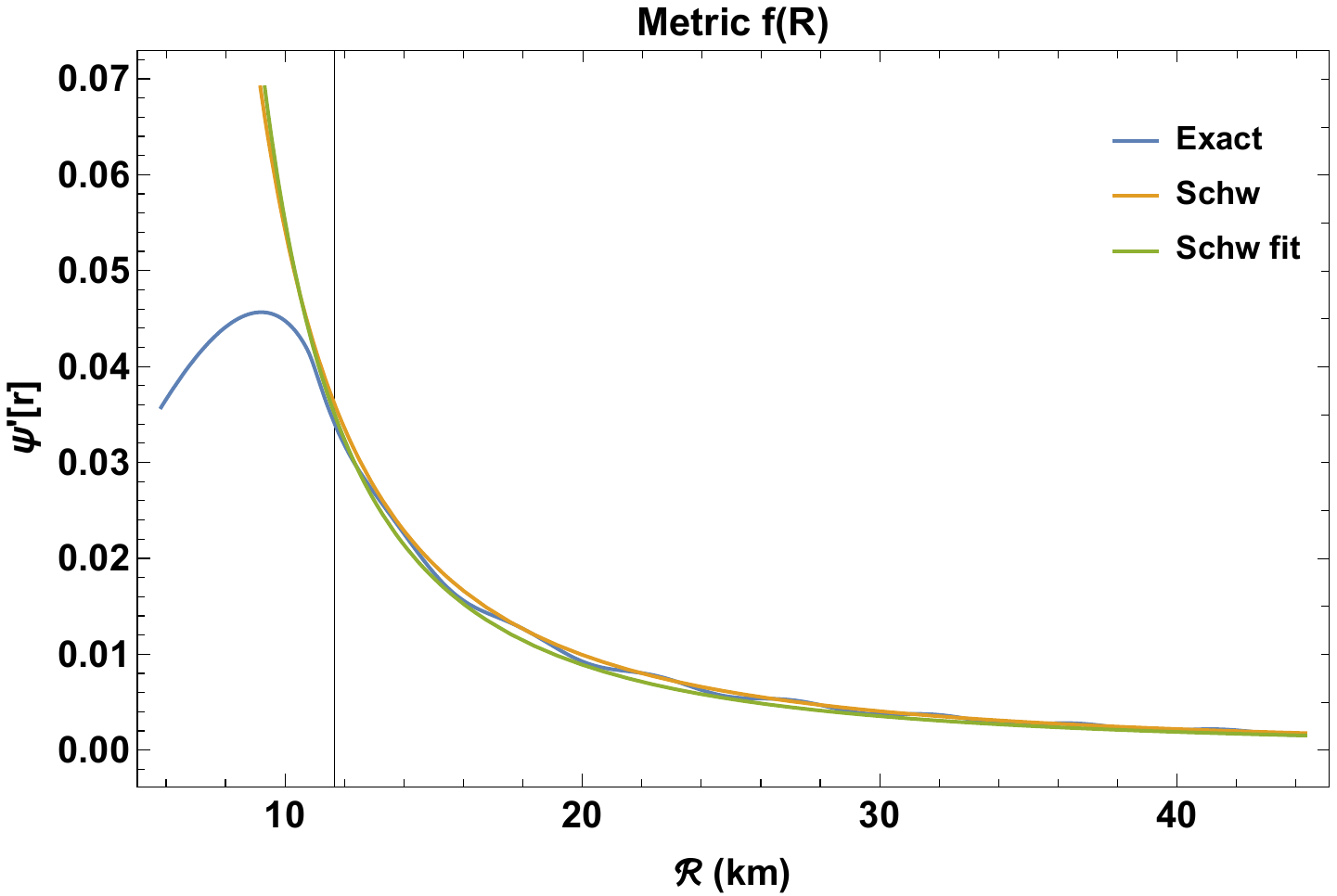}
 \caption{Results of our analysis with $\alpha=0.05$ for $\lambda$ and $\psi$ (left plots) and the derivatives for $R'$ and $\psi'$ (right plots) for the exact numerical solution (blue line); the Schwarzschild solution (orange line) with mass $\mathcal{M}=1.43 {\cal M}_\odot$; a Schwarzschild fit (green line) to the numerical data outside the star, that is, with $\mathcal{R}>11.6 km$. We note that, for $\alpha$ smal and averaging out all the oscillations, all physical quantities reproduce rather well the Schwarzschild solution outside the star, while matching as well the  junction conditions (\ref{junction_metric}). From the fitted results we get ${\cal M}=1.40{\cal M}_\odot$, thus very close to the theoretical one.
  \label{fig:lambapsi}}
\end{figure}

In Fig. (\ref{fig:lambapsi}),  we show the behavior of the metric potentials $\lambda(r)$ and $\psi(r)$ and the derivatives $R'(r)$ and $\psi'(r)$  paying special attention to: (i) the junction conditions at the NS boundary and (ii) their profiles as $r \to \infty$.   We show the full numerical solution (blue line), its  corresponding Schwarzschild solution (orange line) given by eqs. \eqref{eq:lmass} with $M=1.43M_\odot$ and the result of fitting the exterior data to the same Schwarzschild-like \textit{ansatz} in order to quantify the agreement with the Schwarzschild solution outside the star and which results in a NS with total mass $M=1.40M_\odot$.  The good agreement between the three lines confirms that the solution is well approximated by the Schwarzschild solution right outside the star radius  better than $\sim 2\%$. This good match is also extended to their derivatives thus globally satisfying the necessary junction conditions of eqs. \eqref{junction_metric} once the oscillations are averaged out. On the other hand, since the oscillations do not appear on $\psi(r)$, we choose this quantity more appropriated to define the NS mass $\cal M$. 
\begin{figure}[!hb]
 \includegraphics[width=0.48\columnwidth]{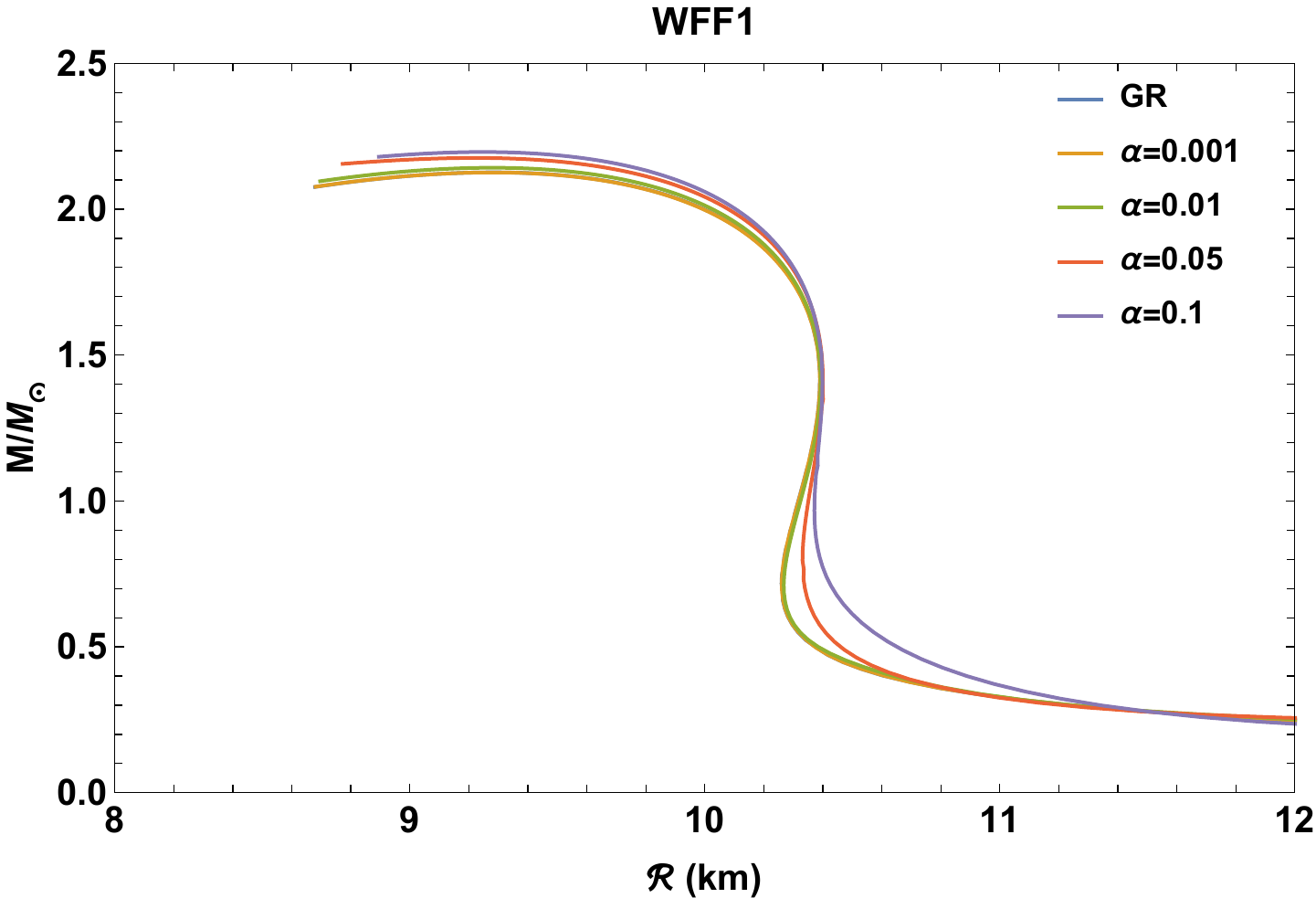}
  \includegraphics[width=0.48\columnwidth]{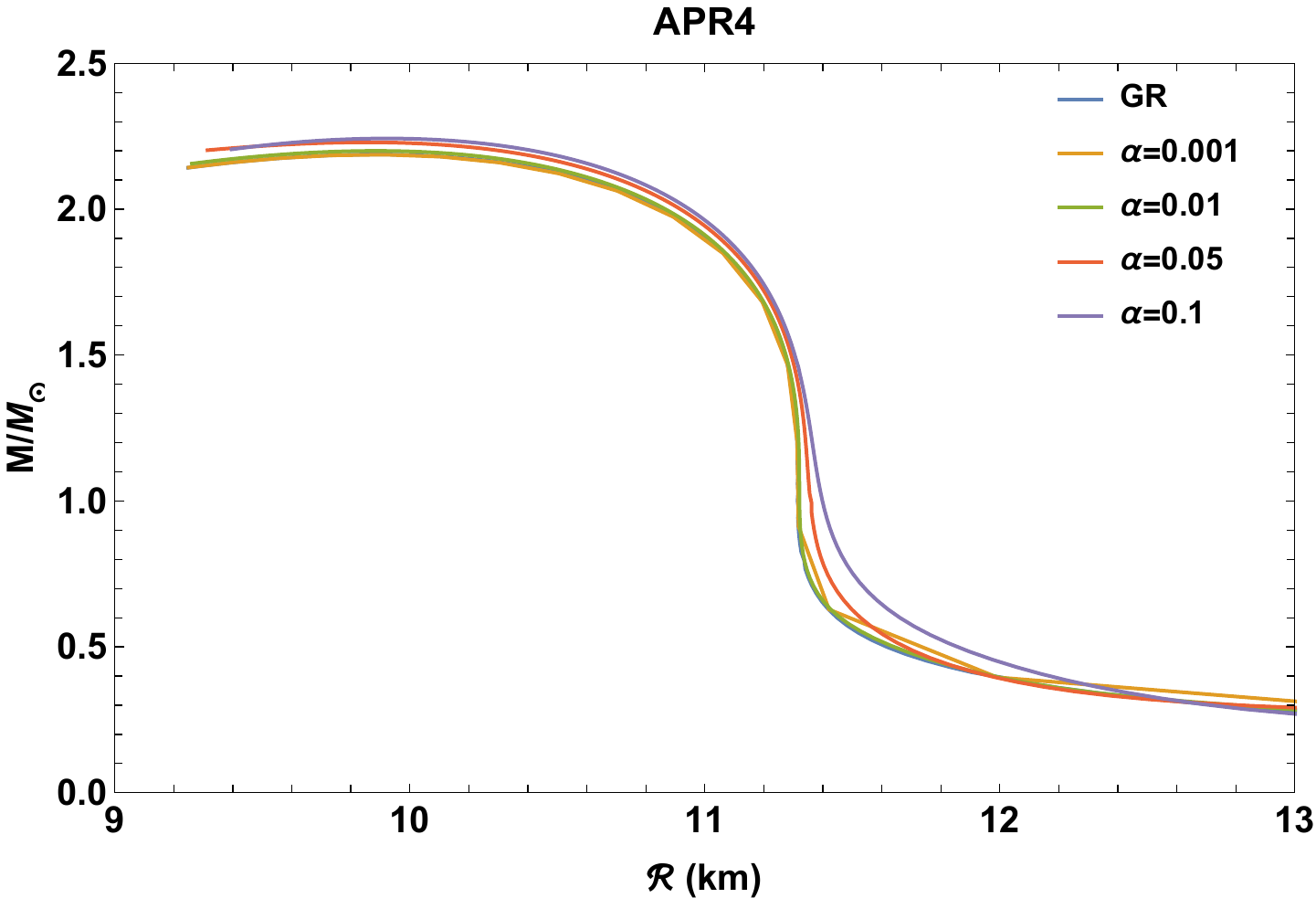}
 \includegraphics[width=0.48\columnwidth]{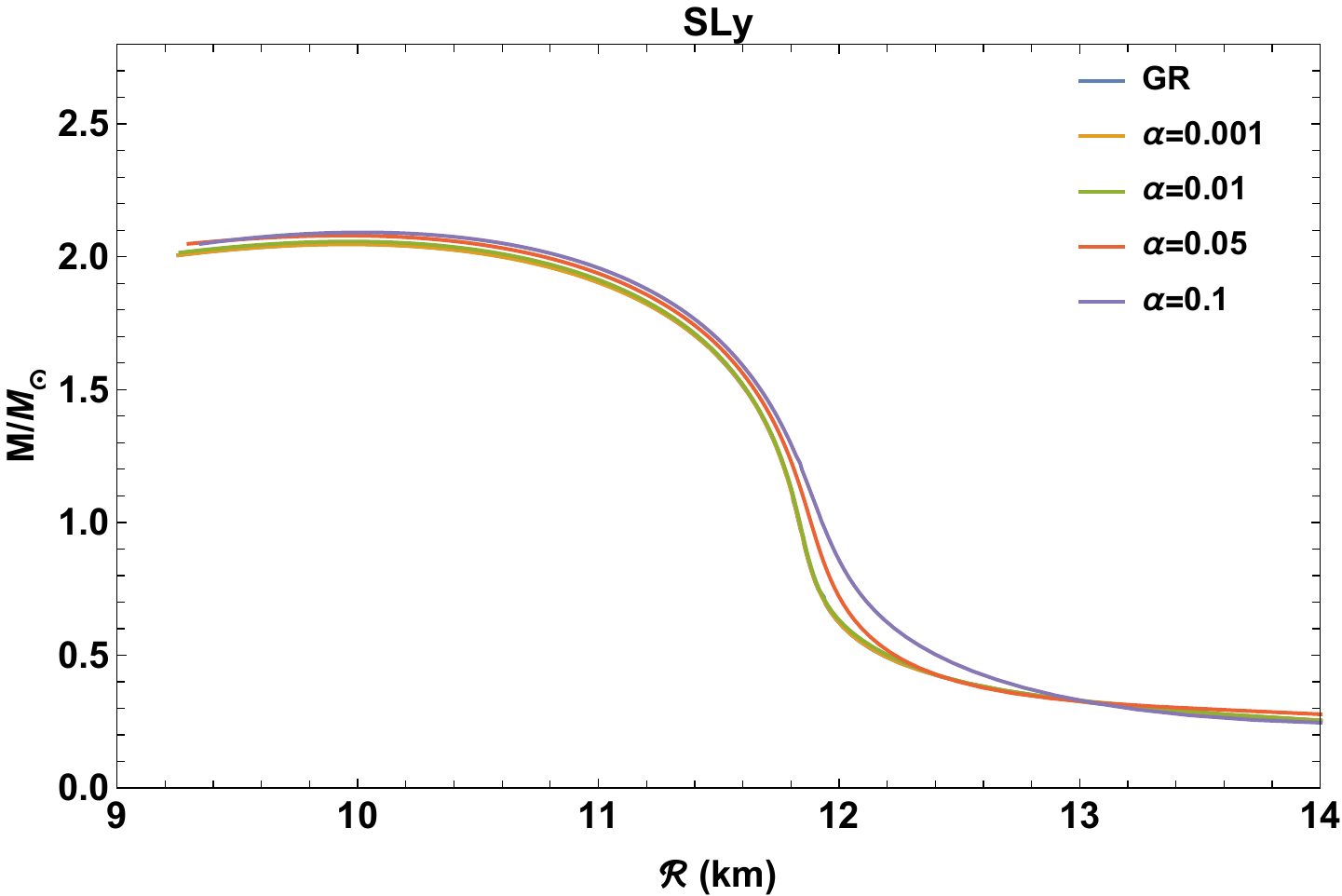}
 \includegraphics[width=0.48\columnwidth]{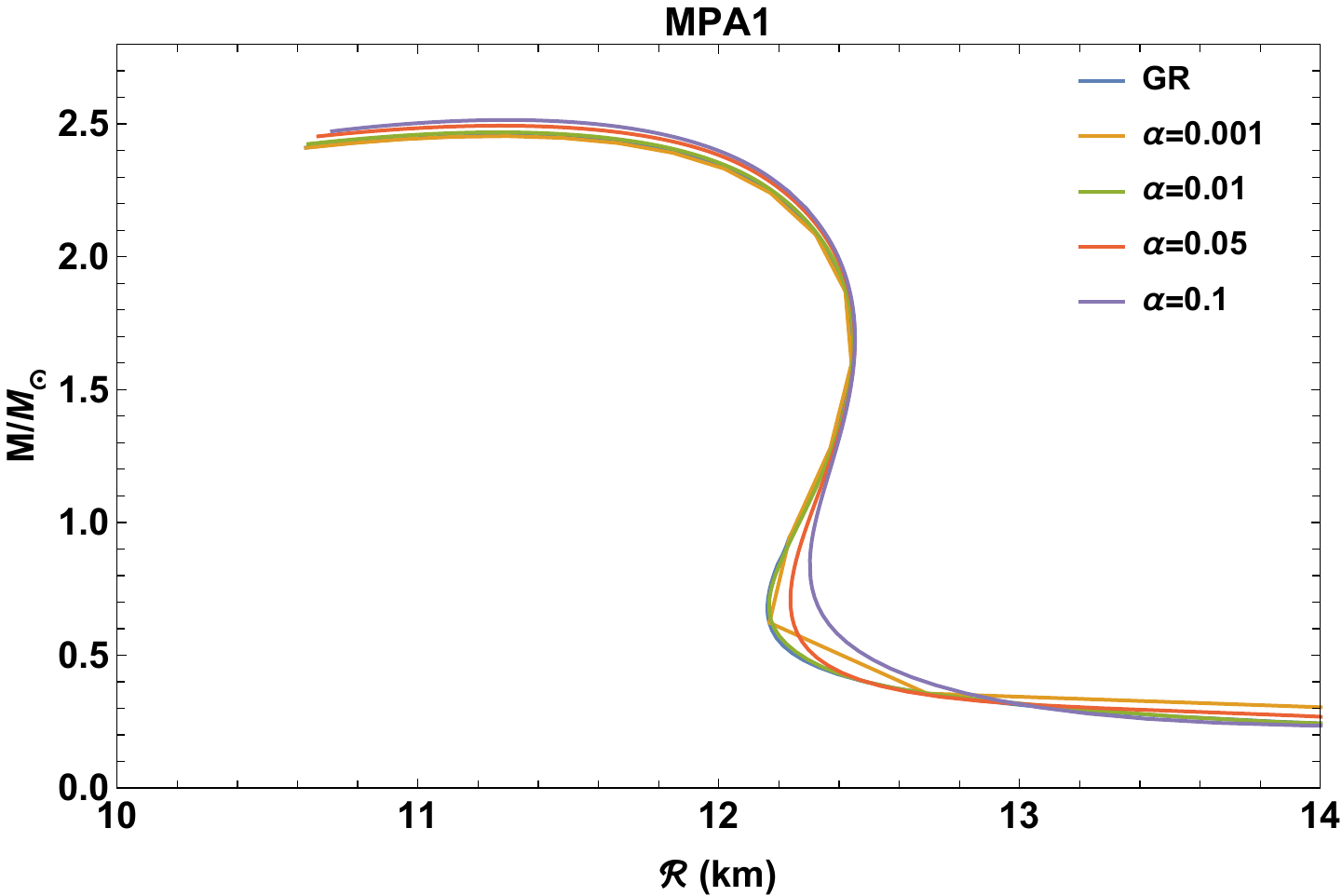}
 \caption{$\mathcal{M-R}$ relations obtained within the purely metric formalism with $\alpha = \left\lbrace 0,0.001,0.01,0.05,0.1\right\rbrace$ for the four EoS considered in this work. Note the general increase of the total mass as the quadratic term takes larger values, thus favoring the formation of more massive objects than in standard GR.
  \label{fig:meMR}}
\end{figure}

Finally, in Fig. \ref{fig:meMR}, we show the $\mathcal{M-R}$ diagrams for the four EoS considered in this work. For each choice of the central density $\rho_c$,  we get a different estimate of the radius $\mathcal{R}_S$ and the total mass $\mathcal{M}$. We loop over $\rho_c$ until $d\mathcal{M}/d\mathcal{R}=0$ which defines the unstable branch, i.e. the point at which the NS is expected to collapse to a black hole and that provides the maximum allowed mass $\mathcal{M}_{max}$ for the given EoS. Note that for all the EoS considered, the total mass tends to increase with respect to GR as in \cite{Doneva:2013rha,Capozziello:2015yza,sbisa}. This is because gravity becomes stronger, thus allowing more massive systems. Indeed, in the $f(R)=R+\alpha\/R^2$ scenario, Newton's gravitational constant $G$ is replaced by
\begin{equation}
\label{eq:Geff}
G \to G_{eff}=\frac{G}{f'(R)}=\frac{G}{1+2 \alpha R}\,.
\end{equation}
The combined conditions of $\alpha >0$ and $R<0$ imply then ${G_{eff}>G}$, thus generating a more \textit{attractive} gravity.
\vspace*{-\baselineskip}
\subsection{Theory with torsion}  
\label{sub:tor}
\begin{figure}[!htb]
 \includegraphics[width=0.48\columnwidth]{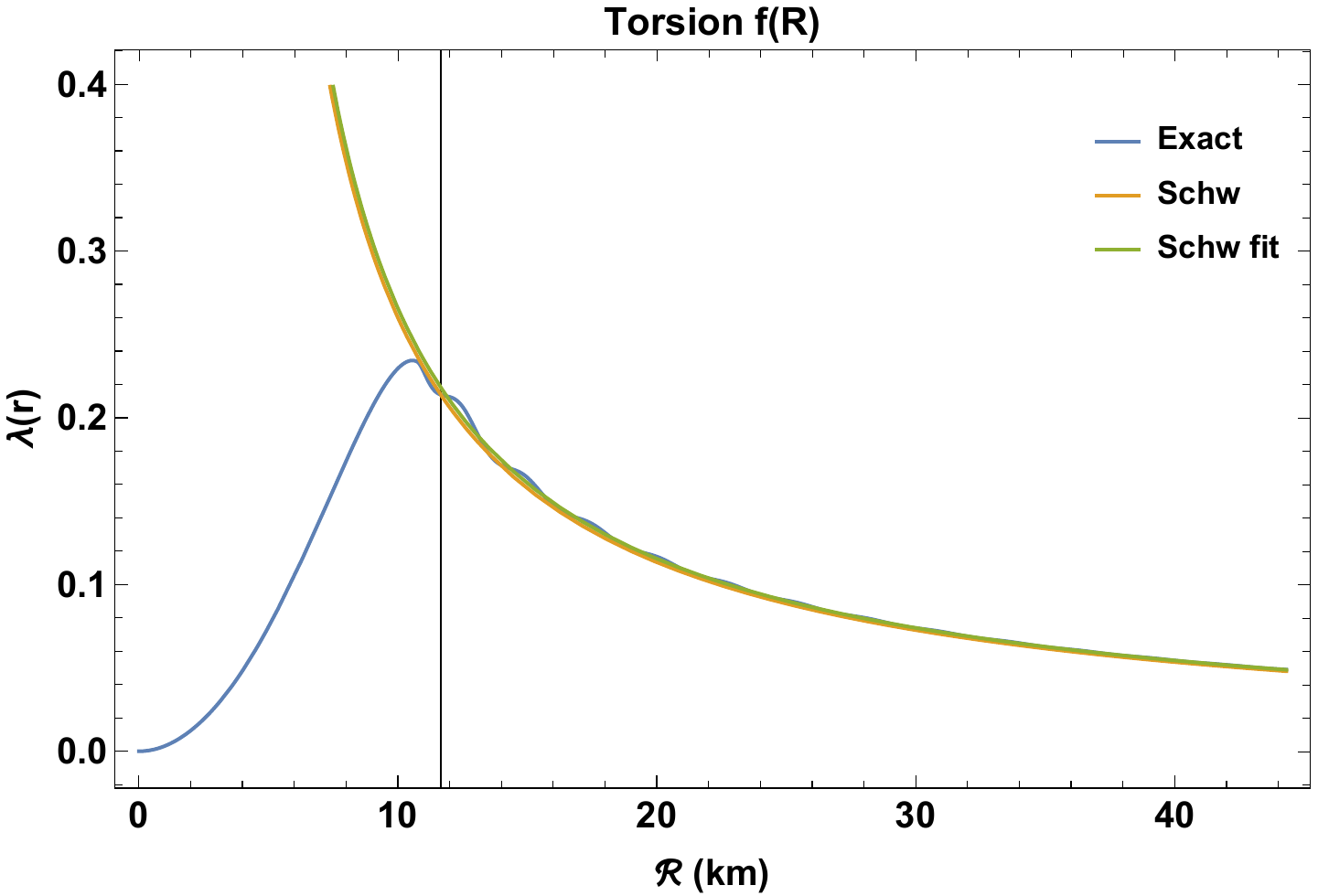}
  \includegraphics[width=0.48\columnwidth]{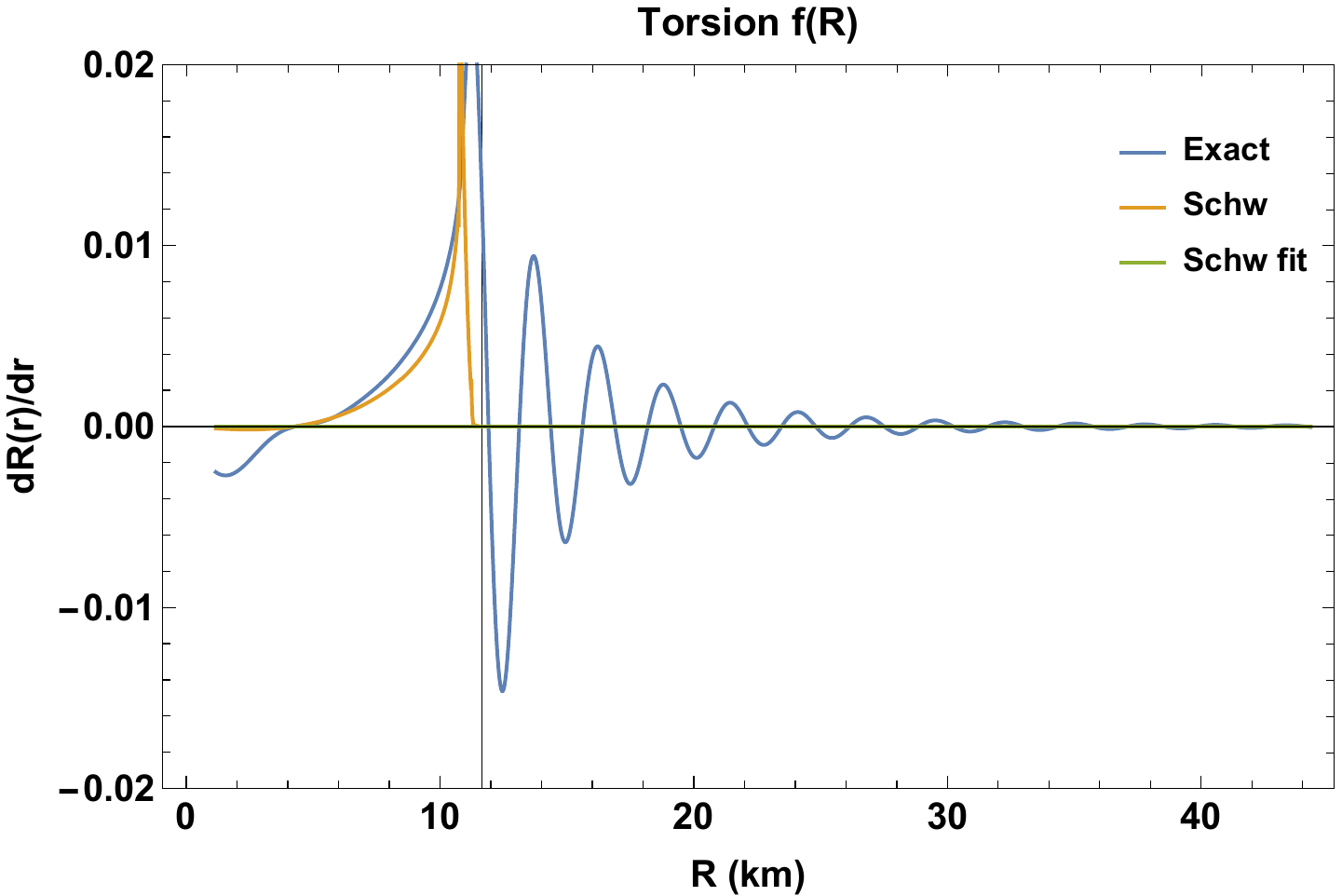}
  \includegraphics[width=0.48\columnwidth]{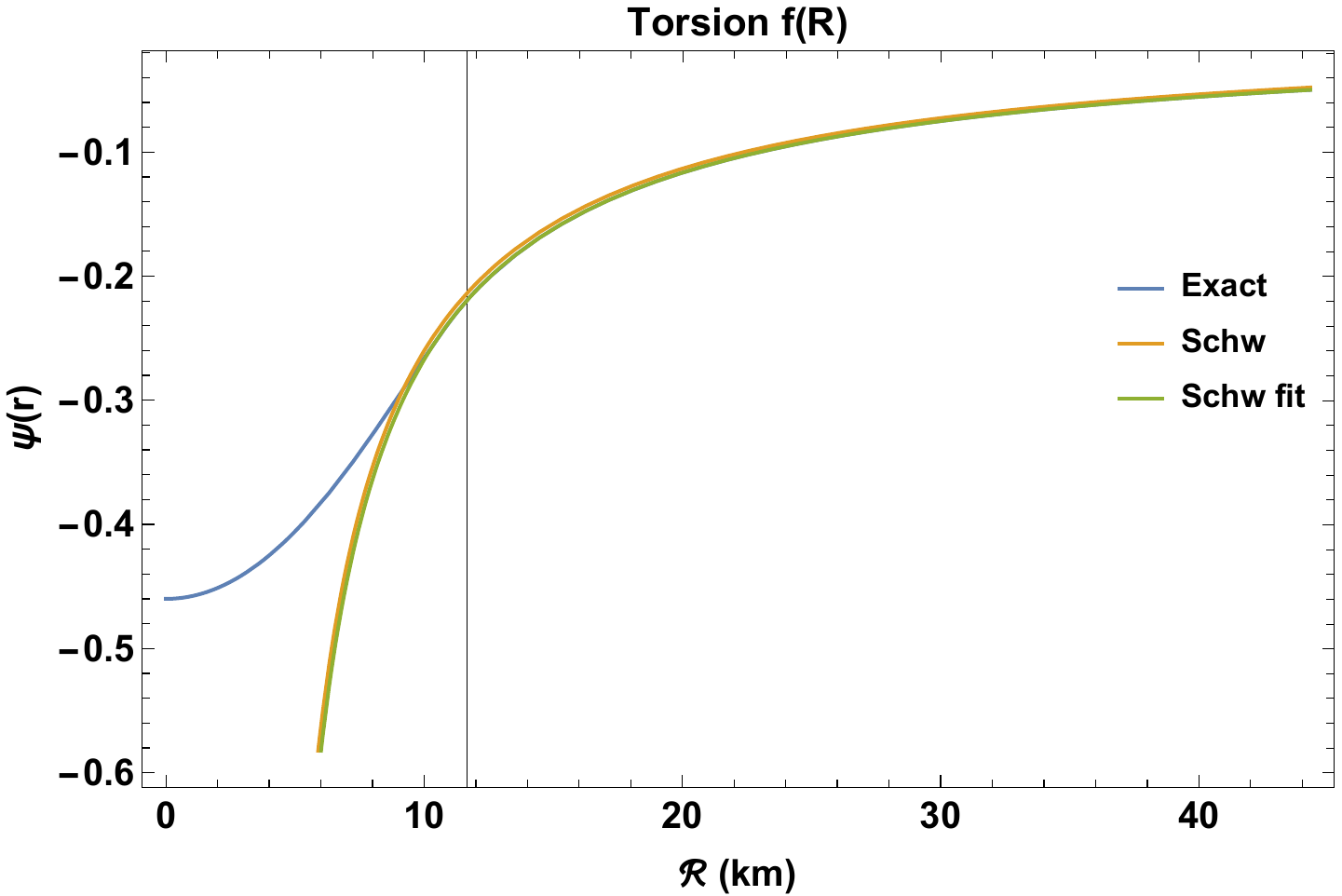}
 \includegraphics[width=0.48\columnwidth]{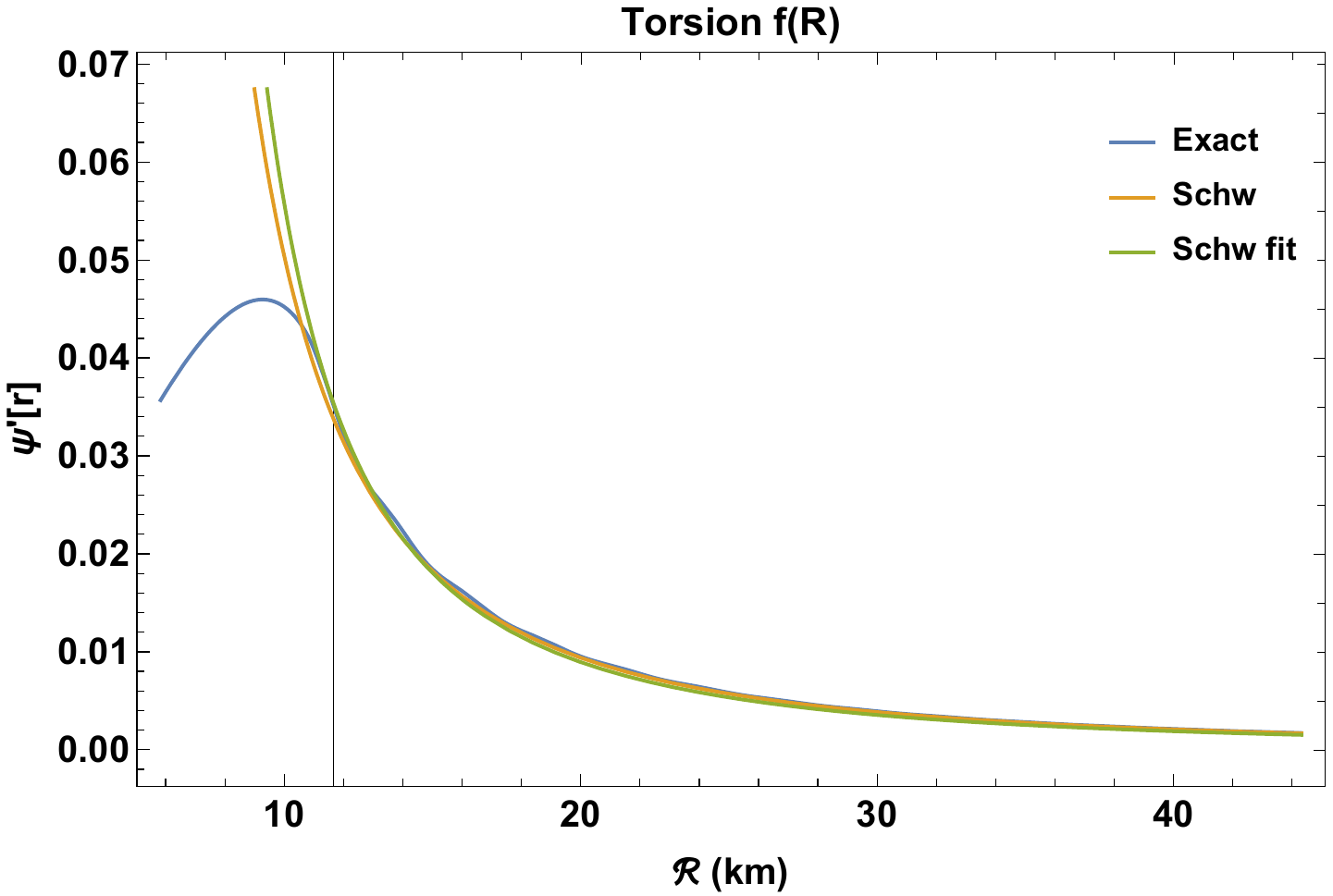}
 \caption{Results of the analysis in the torsional case with $\alpha=0.05$. We show the metric potentials $\lambda$ and $\psi$ (left plots) and the derivatives for $\psi'$ and $R'$ (right plots) for the exact numerical solution (blue line), the Schwarzschild solution (orange line) and a Schwarzschild fit (green line) to the numerical data outside the star, that is with $\mathcal{R}>11.6 km$. Notice that once the oscillations are averaged out, all the distributions satisfy  (up to numerical accuracy) the junction conditions.  
  \label{fig:torlambapsi}}
\end{figure}

\vspace*{-\baselineskip}
We repeat the analysis for the torsional $f(R)=R+\alpha R^2$ theory. Although further models have been also considered in the literature, the numerical complexity of torsional equations makes difficult a full exploration of other kinds of $f(R)$ functions. This issue becomes more relevant when considering the torsional theory with spin  \cite{Capozziello:2008yx}, where spin gradients add higher order derivatives to our system of equations that increase the stiffness of the numerical system. We plan to extend our study in the presence of spin matter in a forthcoming paper.
\begin{figure}[!htb]
 \includegraphics[width=0.48\columnwidth]{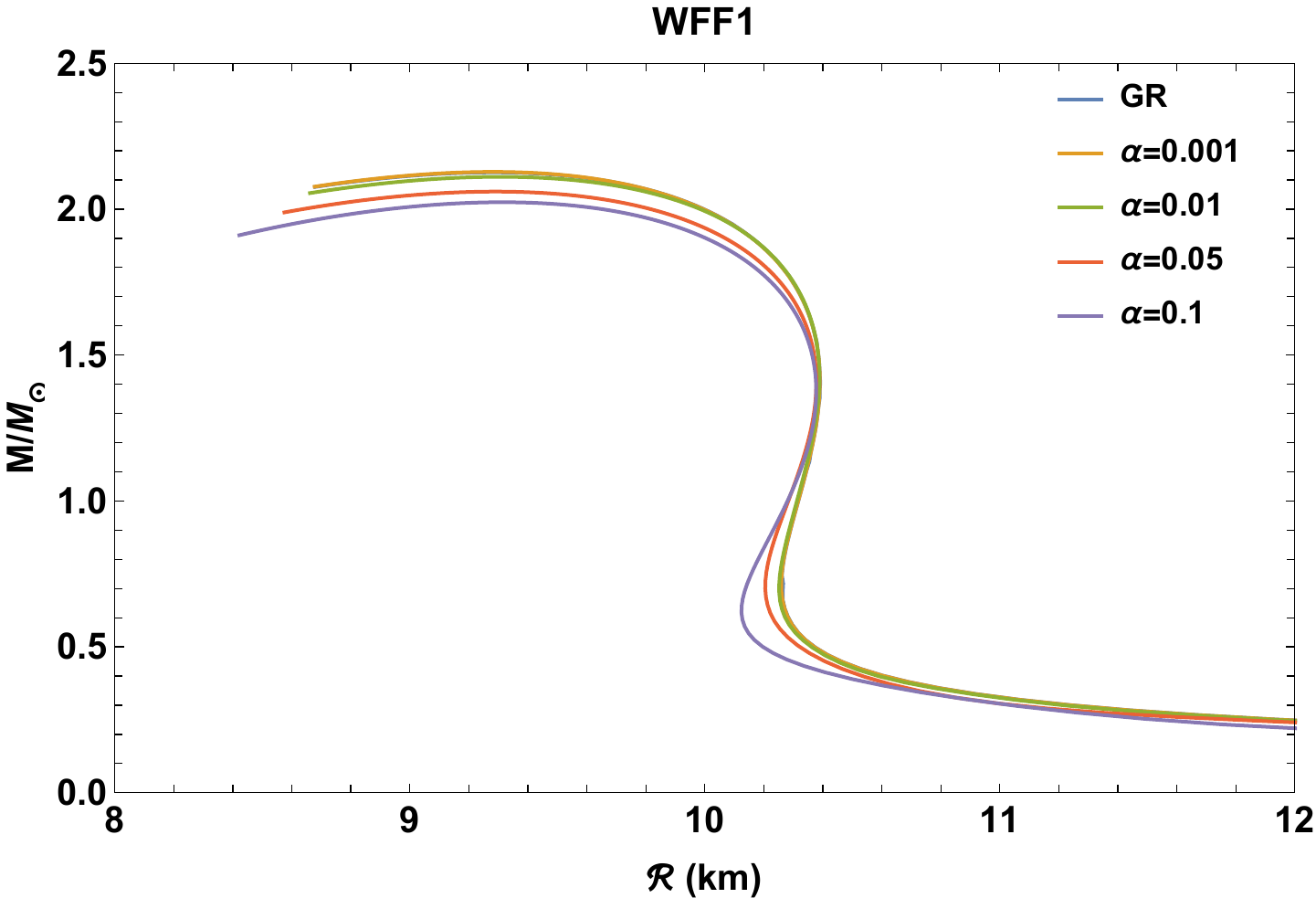}
  \includegraphics[width=0.48\columnwidth]{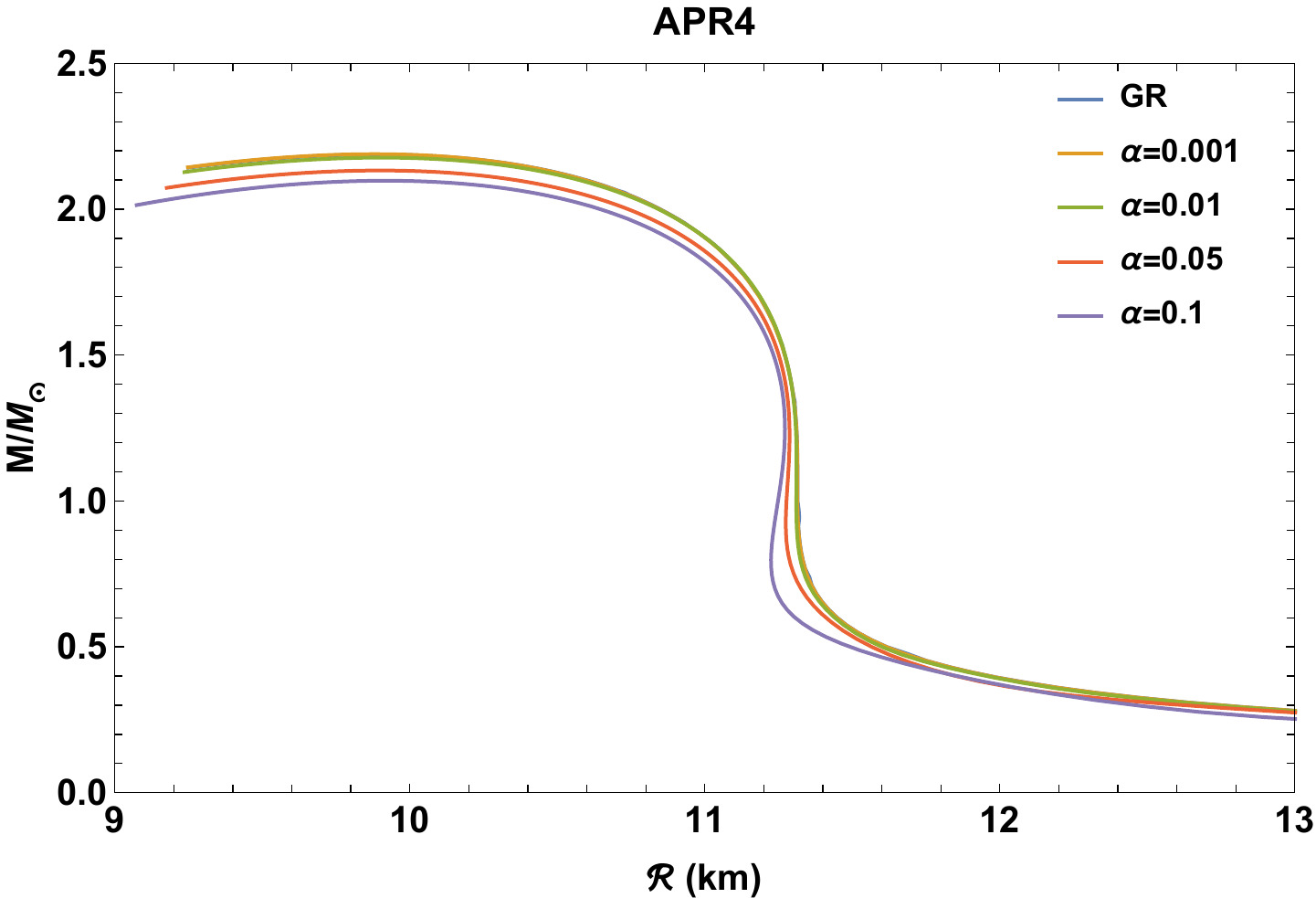}
 \includegraphics[width=0.48\columnwidth]{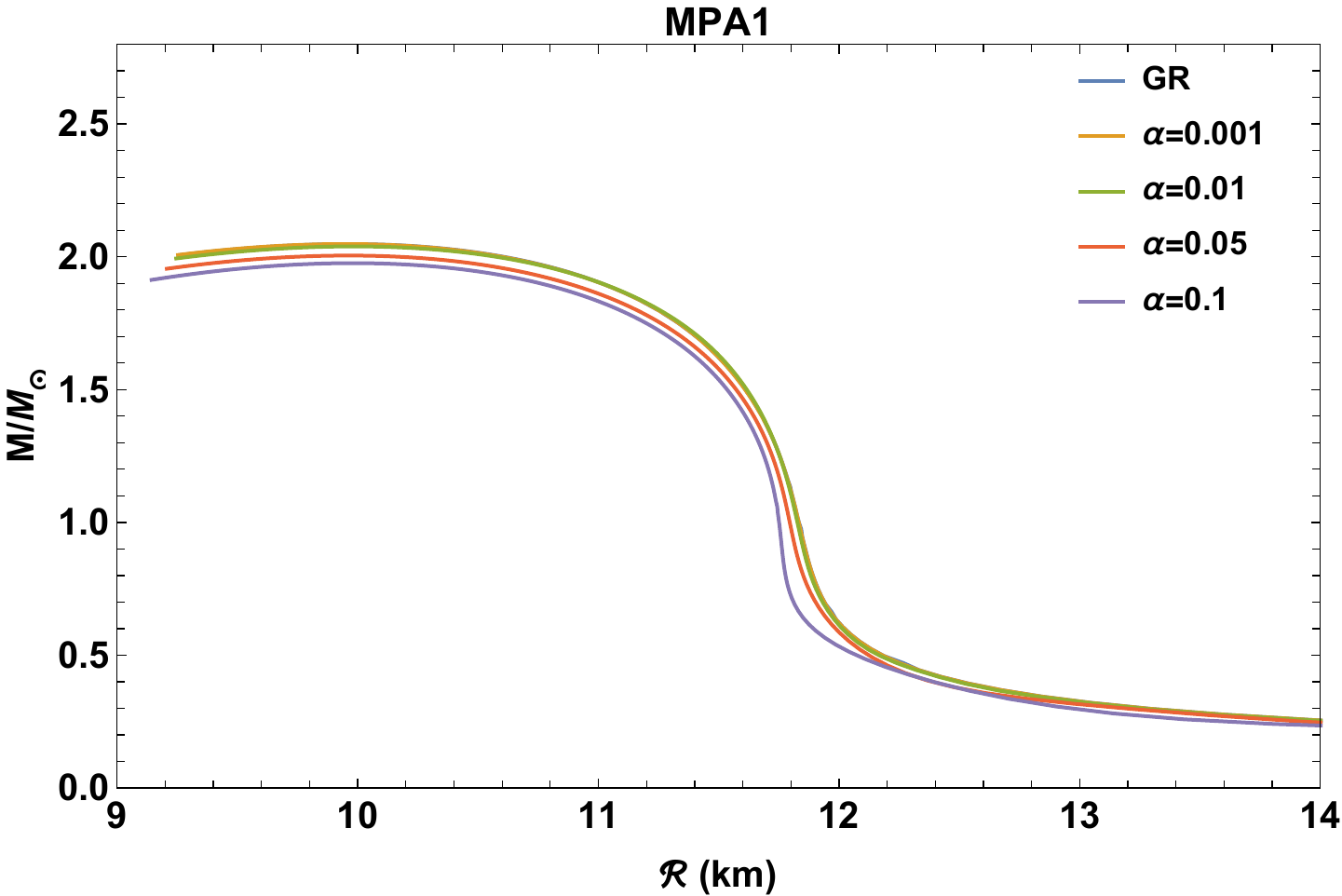}
 \includegraphics[width=0.48\columnwidth]{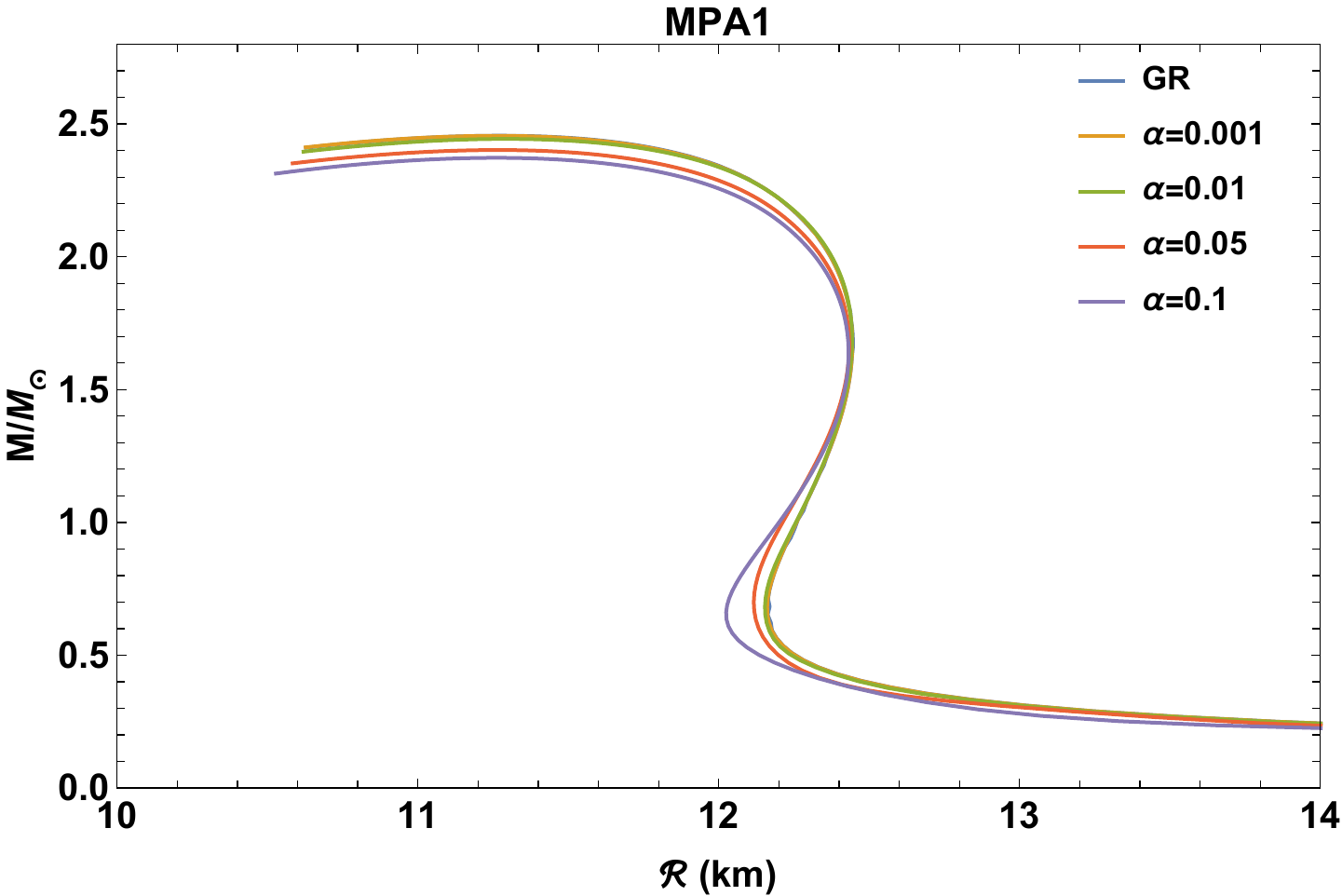}
 \caption{Analogous $\mathcal{M-R}$ relations to those of Fig. \ref{fig:meMR} but here obtained within the torsional formalism. The effect of the torsion tends to decrease the total mass of the NS, contrary to what occurs in the purely metric case. This is dominantly caused by sign flip on the $\alpha$-dependent part of eq. (\ref{TOVR alfa torsion}) with respect to eq. (\ref{TOVR alfa}), which actually acts as a repulsive term.}
  \label{fig:torMR}
\end{figure}
Then, in Figure \ref{fig:torMR}, we show the results we obtained for the theory with torsion, using the same range for $|\alpha|$ as in the metric case but choosing $\alpha<0$. In this scenario, we see that the general trend predicts a decreasing of the total mass of  NS, independently of the EoS considered. This could be related with the fact that the stable branch of the solutions, given by the sign of $\alpha$, is reversed with respect to  the purely metric case to avoid for ghosts. However,  estimates for the total mass and radius are still compatible with the astrophysical observations \cite{Ozel:2016oaf}, thus not allowing us to rule out any of the models studied here. On the other hand, if we further increase $|\alpha|$ the errors generated by eq. (\ref{TOVR alfa torsion}) and propagated to the total mass $\mathcal{M}$ and the total radius $\mathcal{R}_S$ become too large. Therefore, we restrict  our analysis to $|\alpha| \leq 0.1$. In Fig. \ref{fig:torlambapsi} we repeat the same Schwarzschild-based tests adopted  for the metric formalism for $\alpha=0.05$. 
In this case, the total mass ${\cal M}=1.37 {\cal M}_\odot$ is slightly diminished with respect to the metric case.  Notice that the Schwarzschild solution is as well verified at the star radius $\mathcal{R}_s$, where the metric $\lambda(r)$ is clearly $C^0$ and $\psi(r)$ still preserves the $C^1$ condition. Outside the star, and once the oscillations are vanished, the metric functions $\lambda$ and $\psi$ still preserve the $1/r$ decay.
\begin{figure}[!htb]
 \includegraphics[width=0.48\columnwidth]{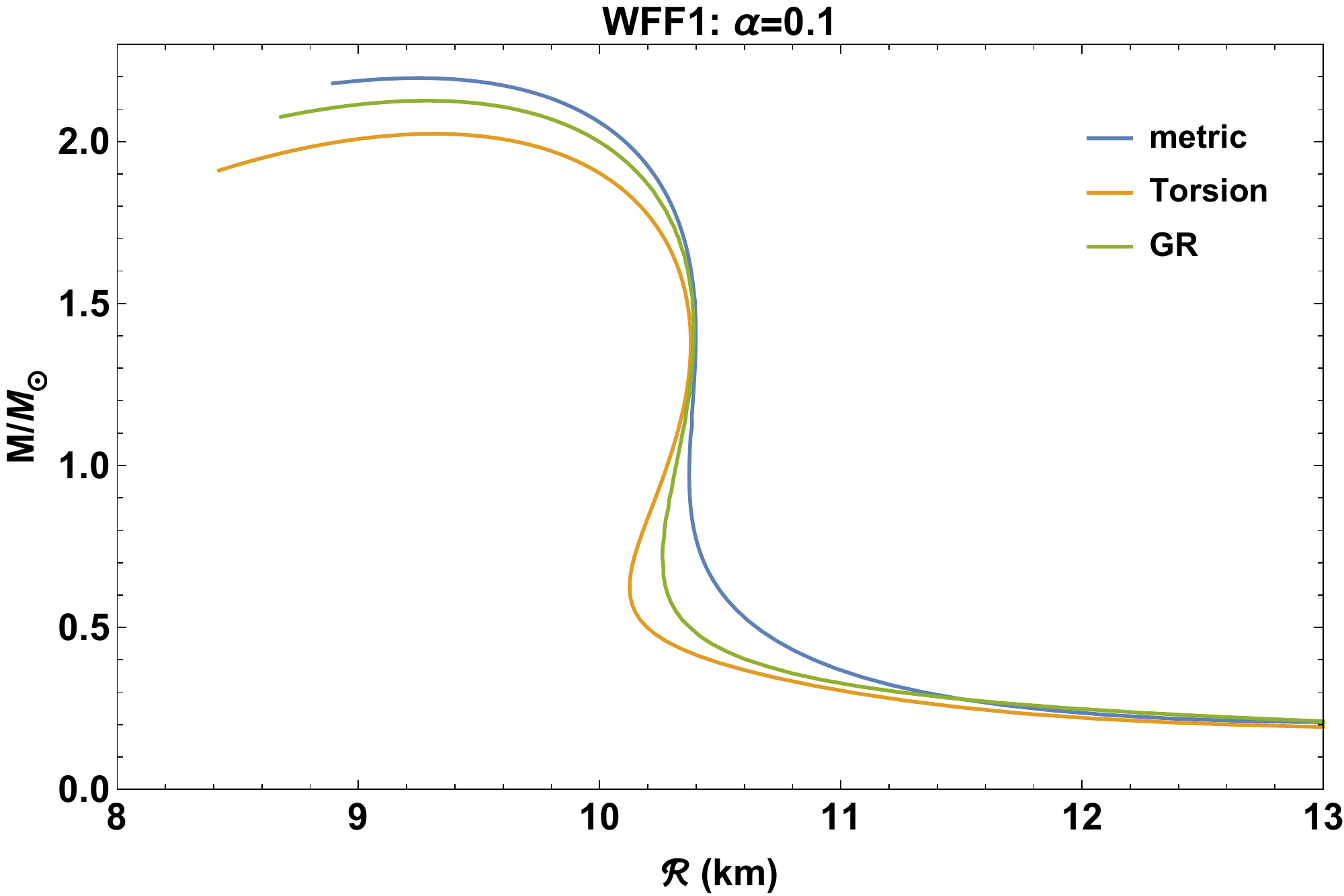}
  \includegraphics[width=0.48\columnwidth]{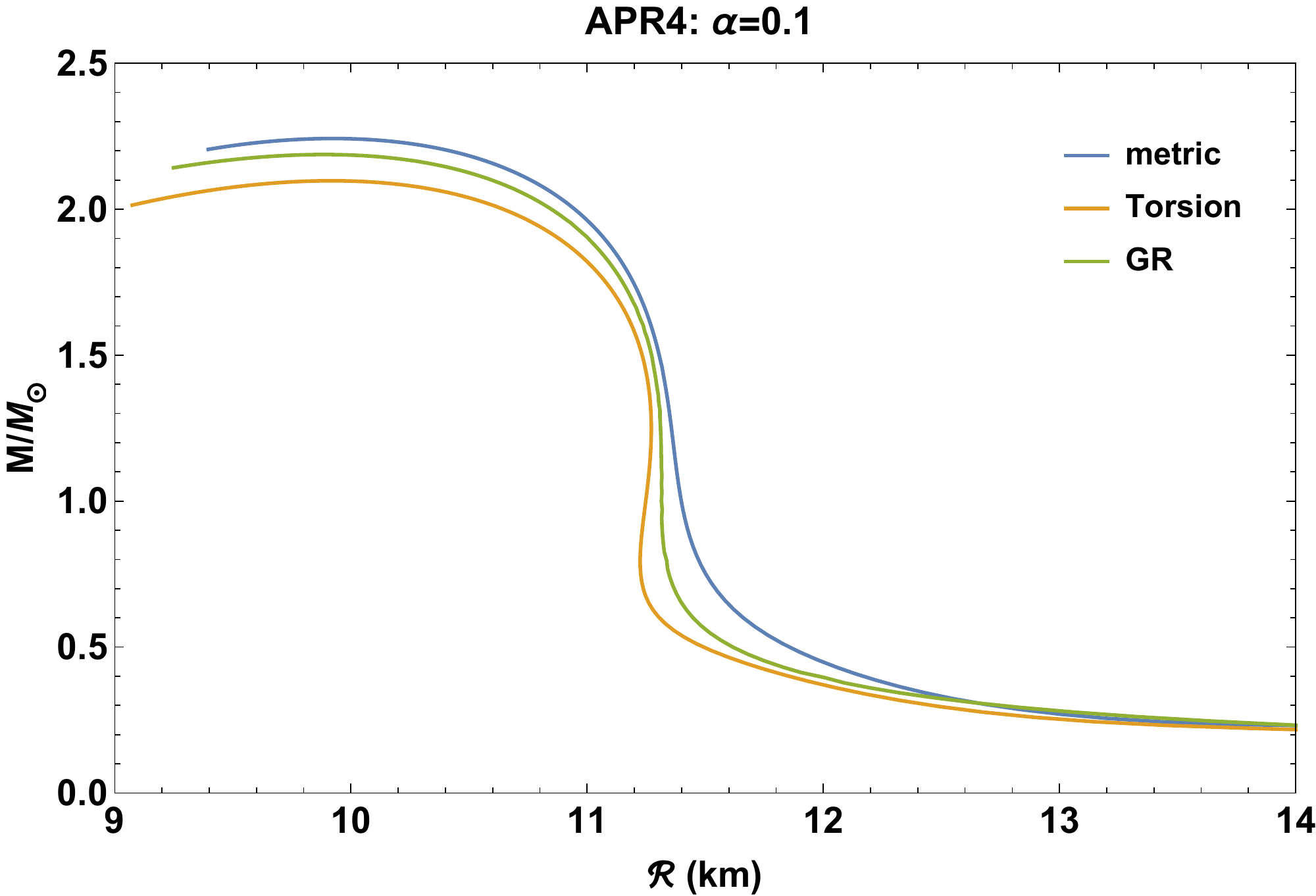}
 \includegraphics[width=0.48\columnwidth]{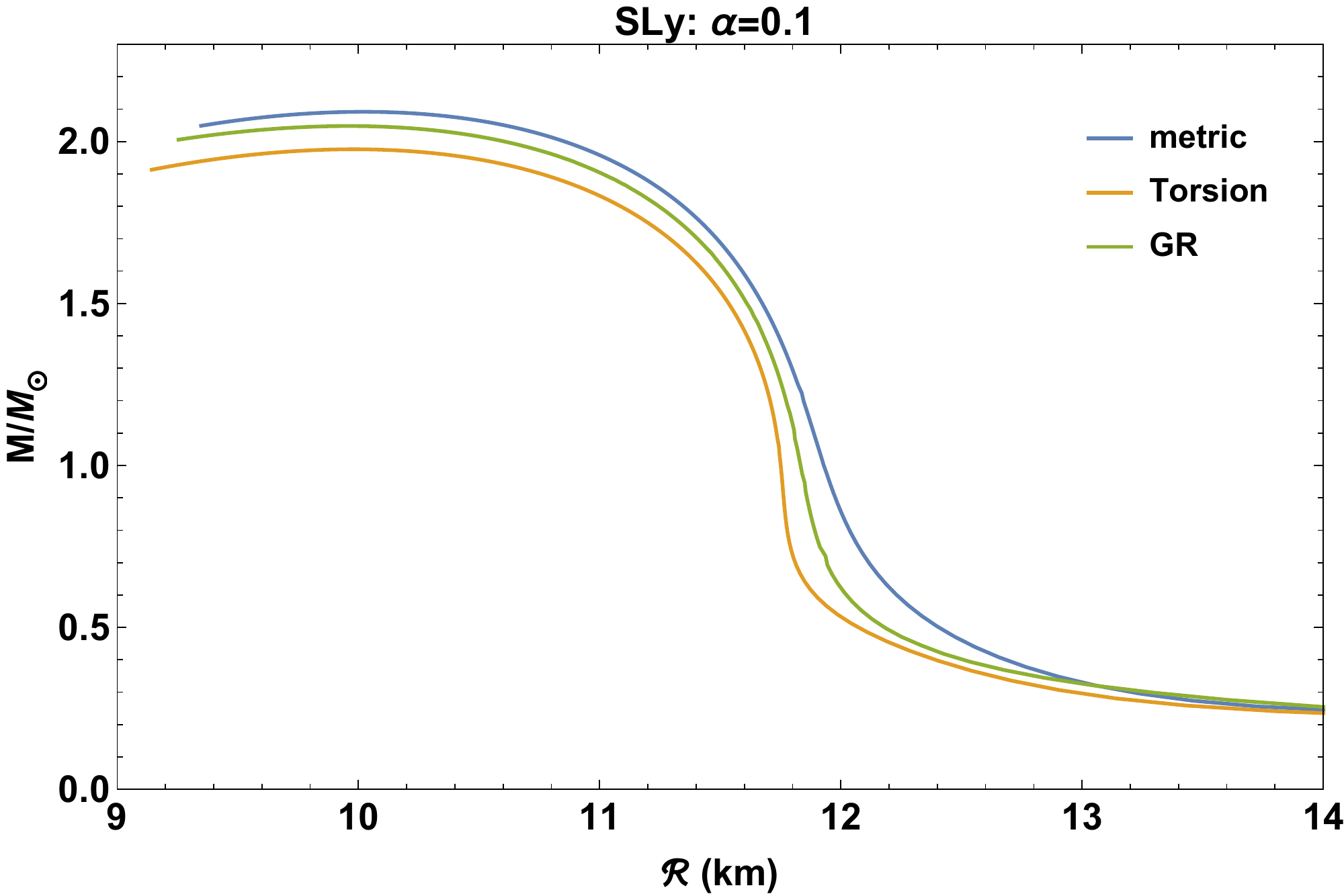}
 \includegraphics[width=0.48\columnwidth]{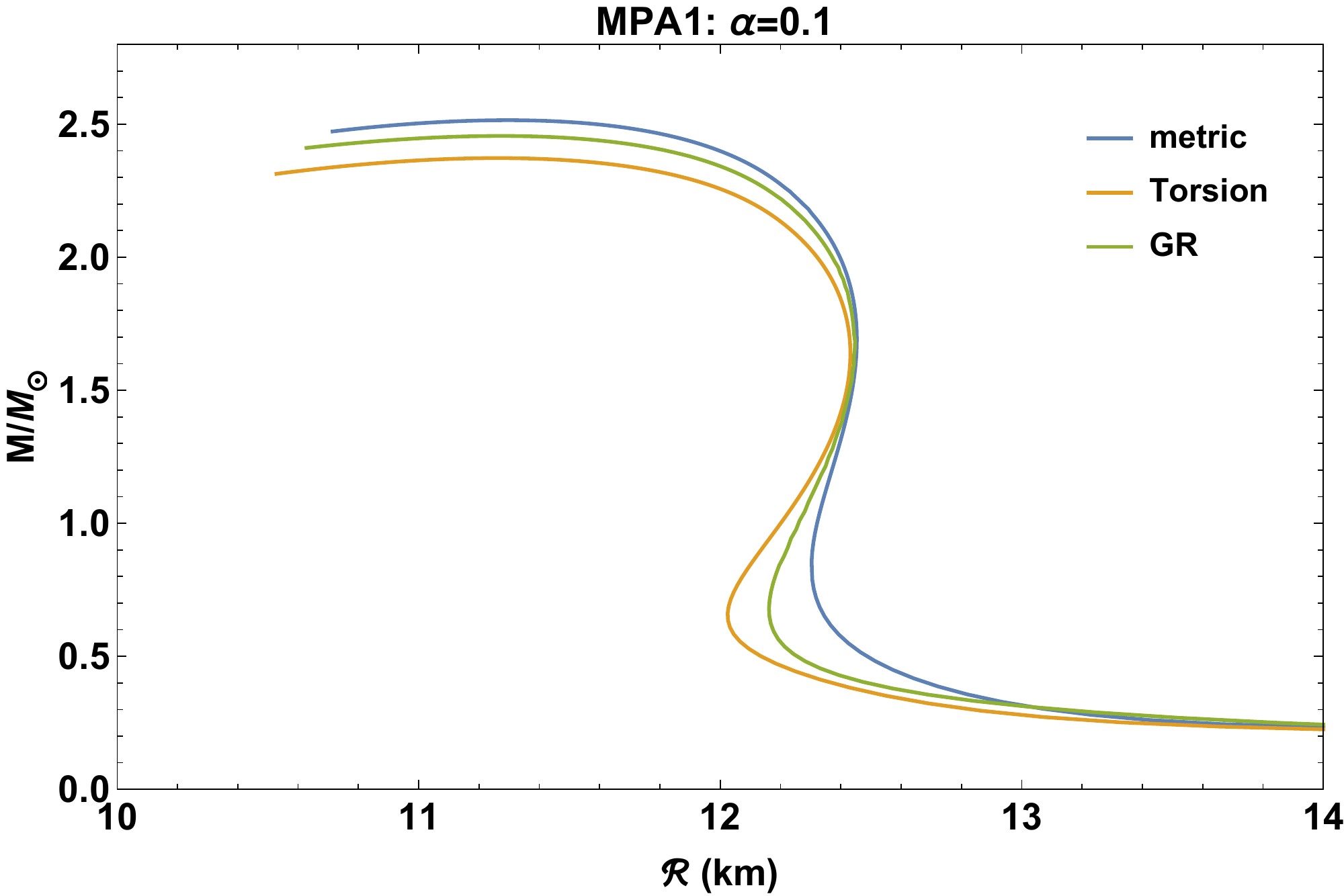}
 \caption{$\mathcal{M-R}$ relations for $\alpha=0.1$ in GR (blue), metric (green) and torsion (orange) for the four EoS considered in this work. The torsion contributions tend to decrease the total mass of the system. 
  \label{fig:torme}}
\end{figure}

Finally, in Fig \ref{fig:torme} we compare the different predictions obtained in the purely metric and the torsional formulation respectively, for $\alpha=0.1$. Note that, in the theory with torsion, though the total mass of the NS decreases, while increases with respect  to the metric case, the relative deviations, in absolute value, with respect to GR seem to be larger than in the metric case. This is caused by the effective repulsion generated by the extra torsional terms (see eq. \eqref{TOVR alfa torsion}) which induce a partial screening of the gravitational field that prevents to reach NS masses as large as in standard GR. This is explicitly shown in Table \ref{tab:NSdata}, where we show the variation of the maximum mass ${\cal M}_{max}$, radius $\mathcal{R}_{max}$ and compactness $C$ for the purely metric and torsional theories respectively, corresponding to the points in the $\mathcal{M-R}$ diagrams where $dM/d\mathcal{R}=0$. Note that whereas the purely metric formulation tends to more massive and compact NSs, the opposite occurs when considering torsion. 
Specifically, as the quadratic term in the curvature increases, the effects of torsion counterbalance the increase of total mass. This can be intuitively derived by using the same reasoning as in \eqref{eq:Geff} with $\alpha <0$ and $R<0$, implying ${G_{eff}<G}$ and thus generating a less \textit{attractive} gravity.

The stability of the solutions can be checked adopting the so-called Regge-Wheeler-Zerilli formalism \cite{Regge,Zerilli}. As discussed  in  \cite{Nashed}, perturbations in $f(R)$ gravity models can be dealt by taking into account odd-type metric perturbations and stability of geodesic motion around the solution. In that  case,  a charged spherically symmetric black hole was considered and stability was strictly dependent on the value of parameters as the black hole  mass, the cosmological constant and the electric charge. In the present case, the leading parameter is $\alpha$ which determines the stability of solution  ${\cal M-R}$.  According to the values reported in Figs \ref{fig:meMR},  \ref{fig:torMR}, and  \ref{fig:torme},  both for metric and torsional case,  our numerical solutions result stable against  perturbations. Specifically, the stability region is given by $d\mathcal{M}/d\mathcal{R}=0$ which  determines the maximal stable configuration as discussed above.   

An important remark is in order at this point to justify the result.
According to the  Regge, Wheeler \cite{Regge}, and Zerilli \cite{Zerilli}  formalism, 
 metric perturbations can be decomposed according to their transformation properties
under two-dimensional rotations.
These authors, originally,   took  into account
 perturbations of the Schwarzschild metric in GR, however, as shown in \cite{Nashed}, the formalism
 depends on the properties of spherical symmetry and then can
be easily applied to $f(R)$ gravity.

If we  consider  the  perturbed metric for a static spherically symmetric spacetime as  $g_{\mu\nu}=g_{\mu\nu}^{0}+h_{\mu\nu}$,   the tensor $h_{\mu\nu}$ represents
small  perturbations with respect to the background. Under two-dimensional
rotations, $h_{tt},h_{tr}$ and $h_{rr}$ transform as scalars, $h_{ta}$
and $h_{ra}$ transform as vectors and $h_{ab}$ transforms as  tensors
($a,b$ are either $\theta$ or $\phi$). A scalar  quantity $\Phi$  is  expressed in spherical harmonics $Y_{\ell m}(\theta,\phi)$ as
 \begin{equation}
\Phi(t,r,\theta,\phi)=\sum_{\ell,m}\Phi_{\ell m}(t,r)Y_{\ell m}(\theta,\varphi).\label{scalar-decomposition}
\end{equation}
Adopting the     spherical symmetry, the solution is independent of   $m$,
and then  this subscript can be omitted and one  can  take into account only the index $\ell$  representing   the multipole  number arising from the separation of angular variables by the
expansion into spherical harmonics, that is 
\begin{equation} \Delta_{\theta, \phi}Y_{\ell}(\theta, \phi)=-\ell(\ell+1)Y_{\ell}(\theta, \phi)\,.
\end{equation}
 A vector $V_{a}$ can be decomposed into a divergence part
and a divergence-free part:
 \begin{equation}
V_{a}(t,r,\theta,\phi)=\nabla_{a}\Phi_{1}+E_{a}^b\nabla_b\Phi_{2},
\end{equation}
where $\Phi_{1}$ and $\Phi_{2}$ are two scalars and $E_{ab}\equiv\sqrt{\det\gamma}~\epsilon_{ab}$.
Here $\gamma_{ab}$ is the two-dimensional metric on the sphere
and $\epsilon_{ab}$ is the totally anti-symmetric tensor with
$\epsilon_{\theta \varphi}=1$;  $\nabla_{a}$ is the covariant derivative with respect to  $\gamma_{ab}$.
Being $V_{a}$  a two-component vector, it
is  assigned by  $\Phi_{1}$ and $\Phi_{2}$. We can
take into account the  decomposition (\ref{scalar-decomposition}) for $\Phi_{1}$
and $\Phi_{2}$ to decompose the vector quantity $V_a$ into spherical
harmonics.
The variables related to  $E_{ab}$ are (axial)
 odd-type modes and the others are  (polar) even-type
modes. This decomposition is useful because, in the linearized equations
of motion (or equivalently, in the second order action) for $h_{\mu\nu}$,
odd-type  and even-type perturbations are completely decoupled. This fact  is due to  the invariance of the background metric  under parity transformations.
Therefore, one can study odd-type  and even-type perturbations
separately. The difference between the two families is their parity.
Under the parity operator $\pi$ a spherical harmonic with index $\ell$ transforms as
$(-1)^\ell$. The polar  perturbations transform, under parity, in the same way.
On the other hand,  the axial perturbations transform as $(-1)^{\ell+1}$. Adopting the Regge-Wheeler formalism,  metric perturbations
are
 \begin{eqnarray}
 &  & h_{tt}=0,~~~h_{tr}=0,~~~h_{rr}=0,\\
 &  & h_{ta}=\sum_{\ell, m}h_{0,\ell m}(t,r)E_{ab}\partial^{b}Y_{\ell m}(\theta,\varphi),\\
 &  & h_{ra}=\sum_{\ell, m}h_{1,\ell m}(t,r)E_{ab}\partial^{b}Y_{\ell m}(\theta,\varphi),\\
 &  & h_{ab}=\frac{1}{2}\sum_{\ell, m}h_{2,\ell m}(t,r)\left[E_{a}^{~c}\nabla_{c}\nabla_{b}Y_{\ell m}(\theta,\varphi)+E_{b}^{~c}\nabla_{c}\nabla_{a}Y_{\ell m}(\theta,\varphi)\right].
\end{eqnarray}
From the gauge transformation $x^{\mu}\to x^{\mu}+\xi^{\mu}$, where $\xi^{\mu}$ are infinitesimal quantities, one can show that some metric perturbations are not physical and then some of them can vanish.  Specifically, we can consider the  transformation:
 \begin{equation}
\xi_{t}=\xi_{r}=0,~~~\xi_{a}=\sum_{\ell m}\Lambda_{\ell m}(t,r)E_{a}^{~b}\nabla_{b}Y_{\ell m},
\end{equation}
where $\Lambda_{\ell m}$  can always set $h_{2,\ell m}$ to vanish.
This is called the Regge-Wheeler gauge. According to this procedure, $\Lambda_{\ell m}$ is
 fixed and there is no remaining gauge degrees of freedom. From this result, the only relevant perturbations are the odd-ones. See also \cite{Ganguly} for details.

\vspace*{-\baselineskip}
\section{Discussion and conclusions}
\vspace*{-\baselineskip}
In this paper, we have studied the existence of realistic NSs in the context of the $f(R) = R + \alpha R^2$ theory both in  purely metric and torsional formulations.  
The main results concern the computation of the $\mathcal{M-R}$ diagrams resulting from the two different theoretical frameworks considered. Matter fields have been represented by static and spherically symmetric perfect fluids where the EoS have been chosen to agree with the recent LIGO-Virgo constraints \cite{TheLIGOScientific:2017qsa}.  The parameter $\alpha$ has been restricted to be smaller than $|\alpha| \leq 0.1 $ to avoid unrealistically large oscillations (see e.g. \cite{Resco:2016upv}) on our metric potentials and therefore ensuring the(i) fullfillment of junction conditions and (ii) the accurate recovery of the Schwarzschild solution far from the source.  These two requirements single out four of the five initial conditions: $p(0)$, $\lambda(0)$,$\psi(0)$ and $R'(0)$, while $R(0)$ remains free. $R(0)$ is ideally defined by choosing this parameter in such a way to match the junction conditions (\ref{junction_metric}),(\ref{junction_torsion}). However, the oscillatory behavior of some solutions for $r \to \infty$ prevents from finding a unique value for $R(0)$. To overcome this issue, we have set $R(0)=R_{GR}$ identical to the GR value. This assumption have been shown to be valid for small $\alpha$, being the estimates of the NS radius only mildly dependent on the $R(0)$ choice, but this is no longer true for $\alpha \gtrsim 1$. 

However, a general consideration is in order at this point to justify the assumption  $R(0)=R_{GR}$.
Let us consider the trace of field equations in metric
\begin{equation}\label{trace1}
f'(R)R-2f(R)+3\Box f'(R)=8 \pi  \Sigma\,,
\end{equation}
and in torsion case
\begin{equation}\label{trace2}
f'(R)R -2f(R)=8 \pi  \Sigma\,.
\end{equation}
Substituting $f(R)=R+\alpha R^2$, we have, in the metric case, 
\begin{equation}
6\Box R-R=8\pi \Sigma\,,
\end{equation}
and, in the torsion case, 
\begin{equation}
R=-8\pi \Sigma\,.
\end{equation}
For the metric picture, it is reasonable to suppose that, at the center of the star, $\Box R \simeq 0$ because one can assume a constant central density without remarkable variations and gradients \cite{kippen}. For the torsion picture, we recover exactly the trace of GR. According to these results, the assumption $R(0)=R_{GR}$, besides the above numerical considerations, is fully justified.


In the purely metric theory, the obtained results show a progressive increasing of the total mass as $|\alpha|$ increases, for all the four EoS considered. This allows for higher masses and more compact NSs than in GR. 
This absloute increasing of the mass and compactness could be also reproduced by assuming softer EoS in GR, consistent with the recent  observations \cite{TheLIGOScientific:2017qsa}. 
In the case with torsion, the NS mass tends to decrease for all the EoS considered. This could be related with the fact that the stable branch of the solutions is flipped with respect to the purely metric case to ensure the stability of the numerical system. The physical existence of such solutions could help us to describe NS compact or not, based on astrophysical observations, choosing the appropriate theory by simply constraining whether $\alpha$ is positive or negative. In the torsional framework, the differences in the $\mathcal{M}-\mathcal{R}$ predictions with respect to GR are larger than those obtained in the purely metric case. As a consequence, the allowed intervals on $\alpha$ are poles apart from the two theories. Moreover, the theory with torsion would seem to describe less compact NS. This would allow one to obtain solutions that could be reproduced using EoS with stiff matter in the limit of GR. Unfortunately, this is in disagreement with the recent LIGO-Virgo discoveries \cite{TheLIGOScientific:2017qsa}. What comes to the rescue is that given the current accuracy of electromagnetic observations, we cannot deny the NS observations yet because the differences with the GR are still too small.  However, this issues could be addressed by next generation gravitational wave detectors (3G)  ~\cite{Sathyaprakash:2012jk,Essick:2017wyl,2017arXiv170200786A}, where  the opportunity to test  results presented in this work could be realistic.

\begin{table}
\begin{center}
 \centering
  \setlength{\tabcolsep}{1em}
\begin{tabular}{|c|c|c|c|c|c|c|c|}
\hline
~~$EoS$ & ~~$|\alpha|$ &~~ $M_{max}^{(M)}$   &~~ $ R_{max}^{(M)}$ &~~ $C^{(M)}$ &~~ $ M_{max}^{(T)}$ &~~ $R_{max}^{(T)}$ &~~ $ C^{(T)}$\\ 
~~~~~    &~~&~~ $M_{\odot}$  &~~ $Km$ &~~ $M_{\odot}/Km$ &~~ $M_{\odot}$ &~~ $Km$  &~~ $M_{\odot}/Km$\\
\hline
    &~~		    0& 2.13 & 9.29 &0.23  & 2.13 & 9.29 & 0.23  \\
	&~~	        0.001& 2.13  & 9.29 & 0.23 & 2.13& 9.29 & 0.23  \\
WWF1&~~         0.01& 2.14& 9.28 & 0.23 & 2.11   & 9.30 & 0.23  \\
	&~~	        0.05& 2.19 & 9.21  & 0.24  & 2.06 & 9.28 & 0.22 \\
	&~~	        0.1& 2.20   & 9.24 & 0.24 & 2.02 & 9.31 & 0.21  \\
\hline
    &~~		    0&2.19  & 9.88 & 0.22 & 2.19  & 9.88 & 0.22  \\
	&~~	        0.001&2.19  & 9.91 &0.22 & 2.19 & 9.88 & 0.22  \\
APR4&~~         0.01&2.20& 9.88 &0.22  &2.18& 9.91 & 0.22 \\
	&~~	        0.05&2.23  & 9.85 &0.23 &2.13 & 9.91 & 0.21  \\
	&~~	        0.1& 2.24    & 9.92  & 0.23 &2.10& 9.91 & 0.21  \\
\hline
    &~~		    0& 2.05  & 9.97 & 0.20 &  2.05  & 9.97 & 0.20 \\
	&~~	        0.001&2.05 & 9.94 & 0.20&  2.05   &  9.94 & 0.20  \\
SLy &~~         0.01&2.06  &9.97&0.21& 2.04& 9.98 & 0.20  \\
	&~~	        0.05&2.08   & 9.94 &0.21 & 2.00& 9.96 & 0.20  \\
	&~~	        0.1&2.10    & 10.02 & 0.21 &1.98 & 9.98 & 0.20  \\
\hline
     &~~		    0& 2.45 & 11.28 & 0.22 &  2.45  & 11.28 & 0.22  \\
	&~~	        0.001&2.45 &11.30 & 0.22 &  2.45& 11.26 & 0.22  \\
MPA1 &~~         0.01&2.47 &11.26 &0.22 & 2.44 & 11.30 & 0.22  \\
	&~~	        0.05&2.50  & 11.28 &0.22 & 2.40 & 11.26 & 0.21  \\
	&~~	        0.1&2.51    & 11.30 & 0.22  & 2.37 & 11.26 & 0.21  \\
\hline
\end{tabular}
\end{center}
\caption{Parameters of Neutron Stars for the EoS  considered in this work for the  $\alpha$ values for the (\ref{fr_form_quadratic}) models in the metric formalism and in a torsion theory. The case $\alpha =0$ is the standard GR. ${\cal M}_{max}$   and  ${\cal R}_{max}$ are the maximum values of mass and radius. The superscripts stand for the $(M)$ metric formalism and $(T)$ torsional formalism, where $C^{(M)}$ and $C^{(T)}$ refer to the compactness ${\cal M}_{max}/{\cal R}_{max}$. 
.}
\label{tab:NSdata}
\end{table}


\section*{Acknowledgements}
We want to thanks Alvaro de la Cruz-Dombriz, Miguel Bezares Figueroa and Carlos Palenzuela for providing useful discussions on the numerical methods used in this work.  SC acknowledges INFN Sez. di Napoli ({\it Iniziative Specifiche} QGSKY and MOONLIGHT2) for support. This article is also based upon work from COST action CA15117 (CANTATA), supported by COST (European Cooperation in Science and Technology). 


\end{document}